# Quantum Machine Learning Protection of Military Quantum Key Distribution Against Cryptographically Camouflaged Attacks

Muhammad Shaheer Bin Junaid
Independent Researcher
Karachi, Pakistan
*contact@mshaheerbjunaid.com*
ORCID: 0009-0002-4088-6232

***Abstract-*** Quantum key distribution proves its protocol secure and says nothing about the hardware beneath it, so military and government operators fielding it for command-and-control keys monitor the channel for implementation attacks, and that monitoring has a blind spot. An adversary with a kleptographic foothold in the generator of a public per-block value $x = g^{v} \bmod p$ can hide attacked blocks in honest noise, gating them on a predicate of its discrete logarithm, making detection a discrete logarithm problem that defeats every efficient classical monitor yet yields to a quantum kernel recovering $v$ through Shor's algorithm. I formalise these cryptographically camouflaged attacks, reduce their hardness to an established learning separation, prove a single-frequency fidelity kernel cannot represent an interval predicate, and test them on Ghillie, a decoy-state BB84 simulator with a positive key rate to 142 km. From 10 to 14 bit groups over two seeds, a classical monitor reads 0.458 to 0.516 on camouflaged attacks while the quantum kernel reads 1.000, and both catch overt attacks above 0.99. Finite-precision recovery under depolarising noise and a hardened predicate lower the quantum result to 0.916 through 0.983 with the classical monitor at chance, and a feasibility probe on IBM Heron processors tracks the exact kernel within 0.034. A defender can therefore discard precisely the compromised key material, although the advantage is asymptotic, awaits fault tolerance, and holds only when the feature map matches the adversary's predicate, since a low-frequency map reads 0.545 on a residue pattern and 0.982 once aligned.

***Keywords-*** Quantum Key Distribution; Quantum Machine Learning; Quantum Kernel Methods; Discrete Logarithm Problem; Military Communications.

## 1. Introduction

Quantum key distribution (QKD) is moving out of the laboratory and onto fielded links, and the organisations with the deepest interest in those links are military ones. NATO approved its first Quantum Technologies Strategy in November 2023 and published a summary in January 2024, committing the Alliance to transition its cryptographic systems to quantum-safe cryptography and to secure communications across the air, space, cyber, land, and maritime domains [1]. China operates the largest deployment in existence, a trusted-relay network spanning more than 10,000 km with 145 fibre backbone nodes, 20 metropolitan networks reaching 80 cities, and six ground stations linked to the Jinan-1 microsatellite [2], built on the earlier integrated space-to-ground network that covered 4,600 km [3]. Command-and-control key material and satellite ground links are the traffic these deployments exist to carry, which is what makes their assurance a military question rather than an academic one.

The threat that matters on such a link is aimed at the equipment, not at the key-exchange proof, which holds under its stated assumptions. A decoy-state BB84 receiver is assembled from real avalanche detectors, real modulators, and real control electronics, and each of those behaves in ways the idealised protocol never describes, so the attacks that exploit the resulting gap are both documented and reproducible. Bright-light blinding drives the detectors out of their single-photon regime and hands over the key while the error rate stays where the operator expects it [4, 5]. The time-shift attack biases which detector fires as a function of arrival time and was demonstrated against a commercial system [6, 7]. Photon-number-splitting and Trojan-horse attacks read what leaks from multi-photon pulses or from light reflected back out of the source optics [8, 9, 10].

Because these attacks live below the protocol, operators fall back on monitoring the link itself, and a link monitor watches per-block quantities such as the sifted gain at each intensity, the quantum bit error rate, the double-click rate, and the arrival-time histogram, then tests for departures from calibrated honest behaviour. That practice is sound on its own terms, and it works well against an attacker who is willing to be loud on the blocks it touches. The whole approach nonetheless rests on a single assumption, which is that an attack leaves a mark somewhere in the monitored quantities. Everything in this paper follows from asking what an operator is left with once that assumption fails.

My concern is the attacker who is patient instead, because nothing obliges an attack to be loud on any individual block, and an eavesdropper who keeps each block's footprint inside the honest sampling envelope leaves no anomaly to find while still degrading the key across many blocks. Under that condition the only thing distinguishing an attacked block from an honest one is the adversary's choice of which blocks to touch. If that choice carries structure, predicting it becomes a learning problem, and the difficulty of the learning problem is a parameter the adversary controls rather than an accident of the channel.

The quantum side of this paper earns its place for a specific reason that the recent literature makes easy to state, because work on quantum machine learning over the last several years has converged on a sober reading, which is that quantum models do not beat classical ones on data without special structure, and that classical learners with access to data match quantum kernels on most tasks anyone has tried [11, 12, 13]. One clean exception has survived that scrutiny and it is the exception this paper builds on, since a quantum kernel has a proven separation over every efficient classical learner whenever the labels are defined through a hard number-theoretic function [14]. My construction places attack detection inside exactly that class, so the property that normally limits quantum machine learning becomes the mechanism that makes the defence work.

The contributions of this paper are the following, listed in the order the sections address them. I define a class of cryptographically camouflaged attacks in which the per-block attack decision is a predicate on the discrete logarithm of a public per-block value, and I give three predicate families of rising difficulty (Section 4). I give the reduction showing that detecting these attacks is hard for any efficient classical monitor under the discrete logarithm assumption and tractable for a quantum kernel whose feature map matches the predicate (Section 4). I prove a small structural result about fidelity kernels which says that a single frequency pair cannot separate an interval predicate at all, because the squared overlap is blind to a shift by half the group, and I confirm it numerically (Section 4.3). I build a decoy-state BB84 telemetry simulator, Ghillie, that generates honest, overt, and camouflaged datasets and reproduces a positive key rate to about 142 km, released together with every benchmark record behind this paper as a permanent archive at https://doi.org/10.5281/zenodo.22906856 [15] (Sections 6 and 7). I show across two seeds and five group sizes that a classical monitor collapses to chance on the camouflaged class once the group is too large to brute force, while the quantum kernel separates it at every size (Section 8.1). I then attack my own result, degrading the exponent recovery to finite precision with device noise and hardening the predicate, and I report where the quantum score falls as a consequence (Section 8.2). I run the overlap circuit that measures a kernel entry on two IBM Heron processors and report the measured kernel against the exact one with full job provenance (Section 9).

I take no position on whether quantum key distribution is the right key-agreement technology for national security systems, and the evidence runs against it. The United States National Security Agency does not recommend it and does not plan to certify such products [16]; a joint position paper from the French, German, Dutch, and Swedish authorities reaches the same conclusion [17]; and the 2026 Post Quantum Cryptography Strategy of the United States Department of War states plainly that quantum key distribution and quantum networking will not be used as a means of achieving security for confidentiality, authentication, or key distribution in its systems [18]. NATO's own summary places post-quantum cryptography as the approach for today and describes quantum key distribution as something that further improvement could later allow to contribute [1]. My work is addressed to the operators who field these links anyway, and China's network shows that they exist at national scale.

## 2. Background

### 2.1 Decoy-state BB84 and its public telemetry

I work with decoy-state BB84 over a lossy fibre channel, using weak coherent pulses at a signal intensity $\mu$, a weak decoy intensity $\nu$, and vacuum [19, 20, 21, 22]. Writing $\alpha$ for the fibre loss in dB/km and $\eta_{\text{Bob}}$ for the receiver efficiency, the transmittance at distance $L$ is

$$\eta(L) = \eta_{\text{Bob}} \cdot 10^{-\alpha L/10}.$$

For a pulse of mean photon number $s$ the overall gain and error rate follow the model of Ma, Qi, Zhao and Lo [22],

$$Q_s = 1 - (1 - Y_0)\, e^{-\eta s}, \qquad E_s Q_s = e_0 Y_0 + e_d \left(1 - e^{-\eta s}\right),$$

with $Y_0$ the background yield, $e_0 = 1/2$ the error rate of a background count, and $e_d$ the optical misalignment error. The single-photon yield, gain, and error rate are

$$Y_1 = Y_0 + \eta - Y_0 \eta, \qquad Q_1 = Y_1\, \mu\, e^{-\mu}, \qquad e_1 = \frac{e_0 Y_0 + e_d\, \eta}{Y_1},$$

and the asymptotic secure key rate per pulse is the Gottesman-Lo-Lütkenhaus-Preskill (GLLP) expression [23]

$$R = q \left[ -Q_\mu f\, H_2 \left(E_\mu\right) + Q_1 \left(1 - H_2 \left(e_1\right)\right) \right],$$

where $q$ is the sifting factor, $f$ the error-correction inefficiency, and $H_2$ the binary entropy. The decoy states let the receiver bound $Y_1$ and $e_1$ from the gains measured at the two intensities [22, 24]. I use the vacuum-plus-weak-decoy bounds for the estimated curve and the exact single-photon values for the reference curve, and I return to finite-key corrections under limitations, since they tighten these bounds and matter for any real deployment [24].

What matters for this paper is the public telemetry rather than the key rate itself. A monitor sees, for each block, the measured gains and error rates at every intensity, the double-click rate, the sifting ratio, and summary statistics of the detection-time histogram. Those quantities carry finite-size sampling noise because a block contains a finite number of pulses, and that sampling spread is the honest envelope. It is also the envelope a camouflaged attack has to hide inside, which is the constraint that shapes everything in Section 4.

### 2.2 Implementation attacks

The four attacks I model differ mainly in how much telemetry footprint they leave, and that difference decides which of them can be camouflaged. Detector blinding is the strongest candidate, because controlling the detectors leaks the key with no added error, and the demonstrated attacks recover raw and sifted key while the quantum bit error rate sits at its honest value [4, 5]. The time-shift attack biases the detector that fires according to arrival time and can be balanced so the bit distribution and the error rate stay undisturbed [6, 7]. Photon-number-splitting shifts the ratio of signal to decoy gains, which is precisely the signature decoy states were designed to expose, making it a poor fit for hiding [8, 22]. Trojan-horse leakage occurs at the source optics and need not touch the monitored channel at all, which makes it invisible to a telemetry-only monitor but also puts its signal outside the data such a monitor collects [9, 10]. The camouflaged datasets in this paper therefore use blinding as the default primitive, and I treat the other two as partial cases. Broader treatments of these attacks and their countermeasures are given by Scarani and colleagues [25], Xu and colleagues [26], Pirandola and colleagues [27], and Lo, Curty and Tamaki [28].

### 2.3 Quantum kernels and the discrete logarithm learning separation

A quantum kernel method maps a classical input $x$ to a state $|\phi(x)\rangle$ and classifies with a support vector machine (SVM) whose kernel is the state-overlap fidelity $K\left(x_i, x_j\right) = |\langle \phi\left(x_i\right)| \phi\left(x_j\right)\rangle|^2$ [11, 12, 29]. Supervised quantum models of this form are kernel methods in the mathematical sense [30], so the question of whether they help reduces to whether their kernel captures structure that no classical kernel reaches efficiently. Kübler, Buchholz and Schölkopf sharpened that question by analysing the spectrum, showing that an advantage is available when the reproducing kernel Hilbert space is low dimensional and aligned with the target function, and that a map spreading its weight over an exponentially large feature space generalises poorly [31]. The construction here sits on the favourable side of that boundary by design, because its useful weight lives in a handful of low frequencies matched to the label.

For data without special structure the answer is that quantum kernels do not obviously help, since classical models with access to data match them on many tasks [13]. The exception this paper is built on is the result of Liu, Arunachalam and Temme [14]. They construct a concept class over the multiplicative group of a prime field, indexed by the discrete logarithm, and prove two things about it. Under the assumption that the discrete logarithm problem is hard, no efficient classical learner classifies the data better than chance by a non-negligible margin; and a quantum kernel whose feature map calls Shor's discrete logarithm routine [32] learns the concept in polynomial time. The quantum feature map is a fault-tolerant circuit, which is why the advantage remains asymptotic and stays beyond what a present-day device can demonstrate at scale. Glick and colleagues generalised the construction to covariant kernels for data with group structure and ran it on a 27-qubit superconducting processor [33], and Agliardi and colleagues have since pushed covariant fidelity kernels to 156 qubits on IBM hardware with an error mitigation strategy built for them [34]. Gyurik and Dunjko then clarified the regime in which the separation is cleanest, which bounds what I claim in Section 4.4 [35].

### 2.4 Military interest in quantum communications

A monitoring gap in quantum key distribution is worth studying because military organisations are building or weighing the links it would affect, even while their own cryptographic authorities argue against the technology. NATO's strategy commits the Alliance to transition its cryptographic systems to quantum-safe cryptography and to research quantum-safe communications across every operational domain, with post-quantum cryptography named as the approach available today and quantum key distribution described as a possible later contributor [1]. RAND has argued that allied militaries need a coherent and interoperable course on the quantum threat instead of an improvised one [36]. China's network is the largest operational example and is presented by its builders as national information-security infrastructure [2]. These deployments are the near-term, trusted-relay end of the longer programme towards a quantum internet set out by Wehner, Elkouss and Hanson [37], and the monitoring question I raise applies to the links being built now rather than to that eventual architecture.

The authorities do not agree with one another here, and that disagreement is itself part of what motivates the work, since the National Security Agency does not recommend quantum key distribution for national security systems and does not plan to certify such products, citing the authentication requirement, the trusted-relay problem, the cost, and the implementation attacks this paper is about [16]. The French, German, Dutch, and Swedish agencies reached a similar conclusion and put the priority on post-quantum cryptography [17]. The Department of War went further in 2026, naming quantum key distribution and quantum networking as technologies that introduce technical security risks and stating that they will not be used for confidentiality, data or entity authentication, key distribution, or randomness generation in its systems, with a migration deadline of 31 December 2031 for post-quantum cryptography across the force [18]. My work sits inside that divide without attempting to settle it, because for the operators who field these links regardless, and China's deployment shows that they exist at national scale, whether a stealthy attack pattern can be detected at all remains a live question, and it is the one I take up here.

## 3. Threat Model

The setting I assume throughout is a point-to-point military link on which two endpoints, Alice and Bob, run decoy-state BB84 over fibre to produce command-and-control key material. A capable adversary reaches the quantum channel and has the patience to work across many blocks instead of forcing a result out of any single one. Figure 1 shows that arrangement together with the data available to each party, and the asymmetry visible there is what the rest of this section makes precise.

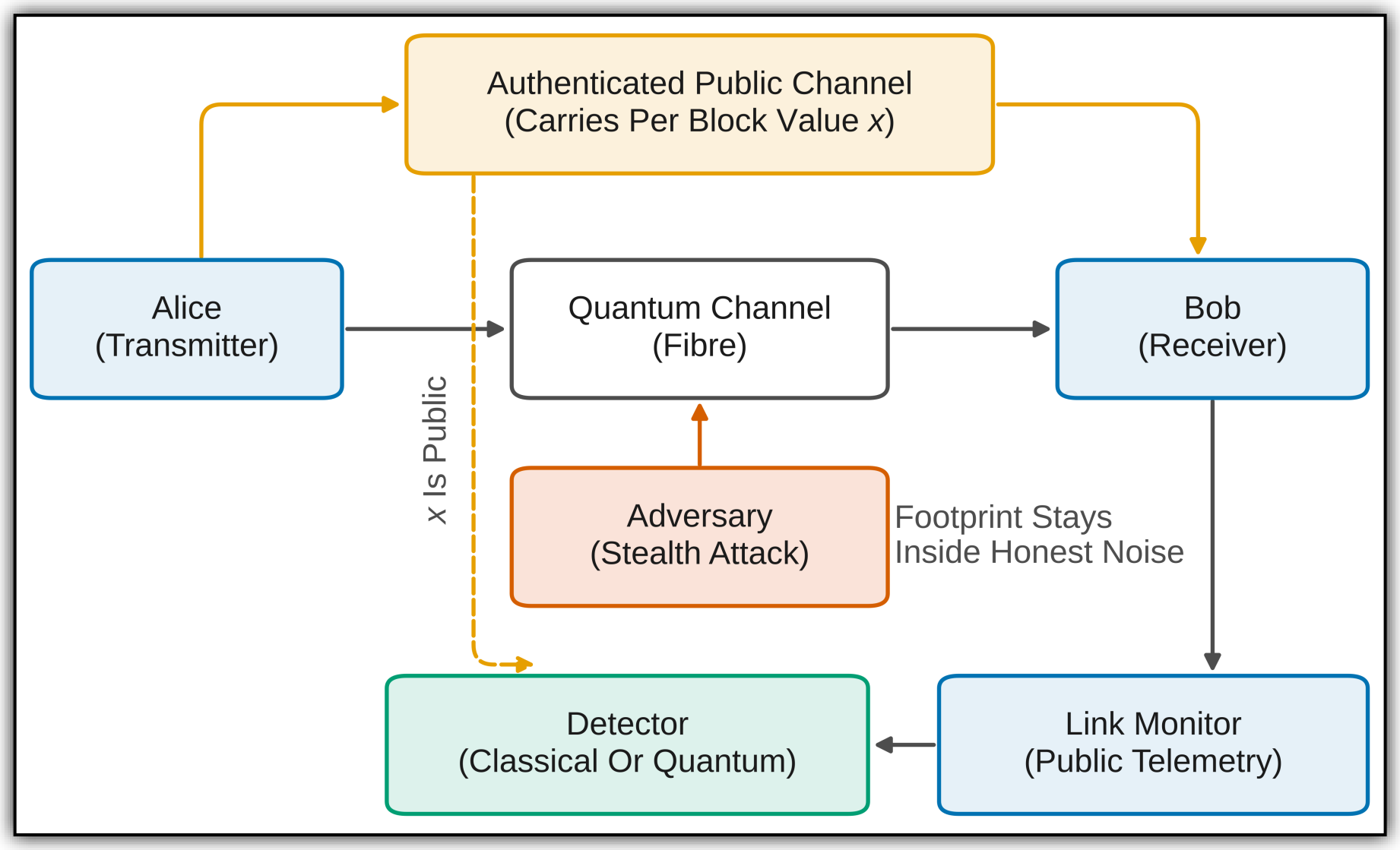

***Fig. 1*** *A deployed decoy-state BB84 link, in which a stealth adversary attacks the quantum channel while the monitoring side sees only the public telemetry and the per-block value x*

Beyond ordinary channel access the adversary needs two further capabilities, of which the first is the ability to tune each attack event so that its telemetry footprint stays inside the honest noise envelope, something blinding supports because it leaks key material without adding error. The second capability is a secret that gates the attack through a function of a public per-block value, which is where the cryptography enters. Both endpoints already exchange such a value over the authenticated public channel, a frame counter or a session nonce being the obvious candidates. The adversary maps that value into the multiplicative group of a prime field as $x = g^v \bmod p$, with $g$ a fixed generator and $v$ the exponent, and attacks a block when $v$ satisfies a chosen predicate. The construction is coherent only where the adversary controls the component that produces the public value, and the natural way to arrange that is a kleptographic implant in the generator, so that the group element travels on the wire as ordinary protocol data while the exponent never leaves the party who planted it [38, 39]. Young and Yung showed that discrete-logarithm cryptosystems admit precisely this kind of secretly embedded trapdoor, and the Dual_EC_DRBG generator standardised by NIST is the documented case of such a trapdoor reaching certified, deployed implementations [40]. That assumption is load-bearing in both directions, because a nonce-to-exponent map the monitor could recompute would hand it the label outright, while a map opaque enough to hide the exponent from the monitor would hide it from the adversary as well. Having built the construction, the adversary knows $v$; any observer handed $x$ must solve the discrete logarithm to recover it.

The observables available to the monitor are the per-block telemetry of Section 2.1 together with the public value $x$, and the asymmetry is that the monitor sees $x$ but never $v$. The defender's goal is not to certify the channel clean, which under camouflage is unachievable from telemetry alone. It is to predict which blocks were attacked, so the key material from those blocks can be discarded or given heavier privacy amplification. Detection in this setting means predicting a hidden structured pattern rather than finding an anomaly, because under camouflage there is no per-block anomaly to find.

Two things sit outside my scope and I state them plainly instead of leaving them implied. I do not defend against attacks on the key-exchange proof itself, which is a separate problem with its own literature [25, 26]. And I do not argue that quantum key distribution should be used for national security systems, since the authorities that advise against it do so on the strength of the same implementation attacks I model here [16, 17, 18].

## 4. Cryptographically Camouflaged Attacks

### 4.1 The camouflaged-attack class

Fix a prime $p$, a generator $g$ of the multiplicative group $\mathbb{Z}_p^*$, and let $M = p - 1$ be the order of the exponent group, which is the setting Diffie and Hellman introduced for key agreement and the one whose hardness the whole construction borrows [41]. For each block let the public value be $x = g^v \bmod p$ with $v$ drawn uniformly from $\{1, \ldots, M\}$. A camouflaged attack pairs a stealth attack primitive with a predicate $a : \mathbb{Z}_p^* \to \{0, 1\}$ evaluated on the discrete logarithm, so that the block is attacked when $a(x) = 1$ and left honest otherwise, with the attacked footprint tuned to stay inside the honest envelope.

I study three predicate families, ordered by how hard they are for a detector to represent. The first is the single interval used by Liu, Arunachalam and Temme [14],

$$a_{\text{int}}(x) = \mathbb{1}\left[\log_g x \in [1, M/2]\right].$$

The second is a union of $m$ disjoint bands spread through the group, which no single frequency describes,

$$a_{\text{band}}(x) = \mathbb{1}\left[\log_g x \in \bigcup_{i=1}^{m} [\ell_i, r_i]\right], \qquad [\ell_i, r_i] \cap \left[\ell_j, r_j\right] = \varnothing \;\text{ for } i \neq j.$$

The third is a residue pattern keyed on the exponent modulo two small primes $q_1$ and $q_2$, in the manner of the Chinese remainder theorem,

$$a_{\text{res}}(x) = \mathbb{1}\left[(v \equiv 0 \bmod q_1) \oplus (v \equiv 0 \bmod q_2)\right], \qquad v = \log_g x,$$

with $\oplus$ the exclusive or. Every experiment below fixes three disjoint bands for the union family and the two smallest odd primes, three and five, for the residue family. The residue family matters because its structure lives at frequencies tied to $q_1$ and $q_2$ rather than at the low end of the spectrum, so a feature map built only from low frequencies cannot see it at all. That turns out to be one of the more useful findings in the paper, and I return to it in Section 8.3. Figure 2 shows the gate in operation, from the public nonce through to the attack decision.

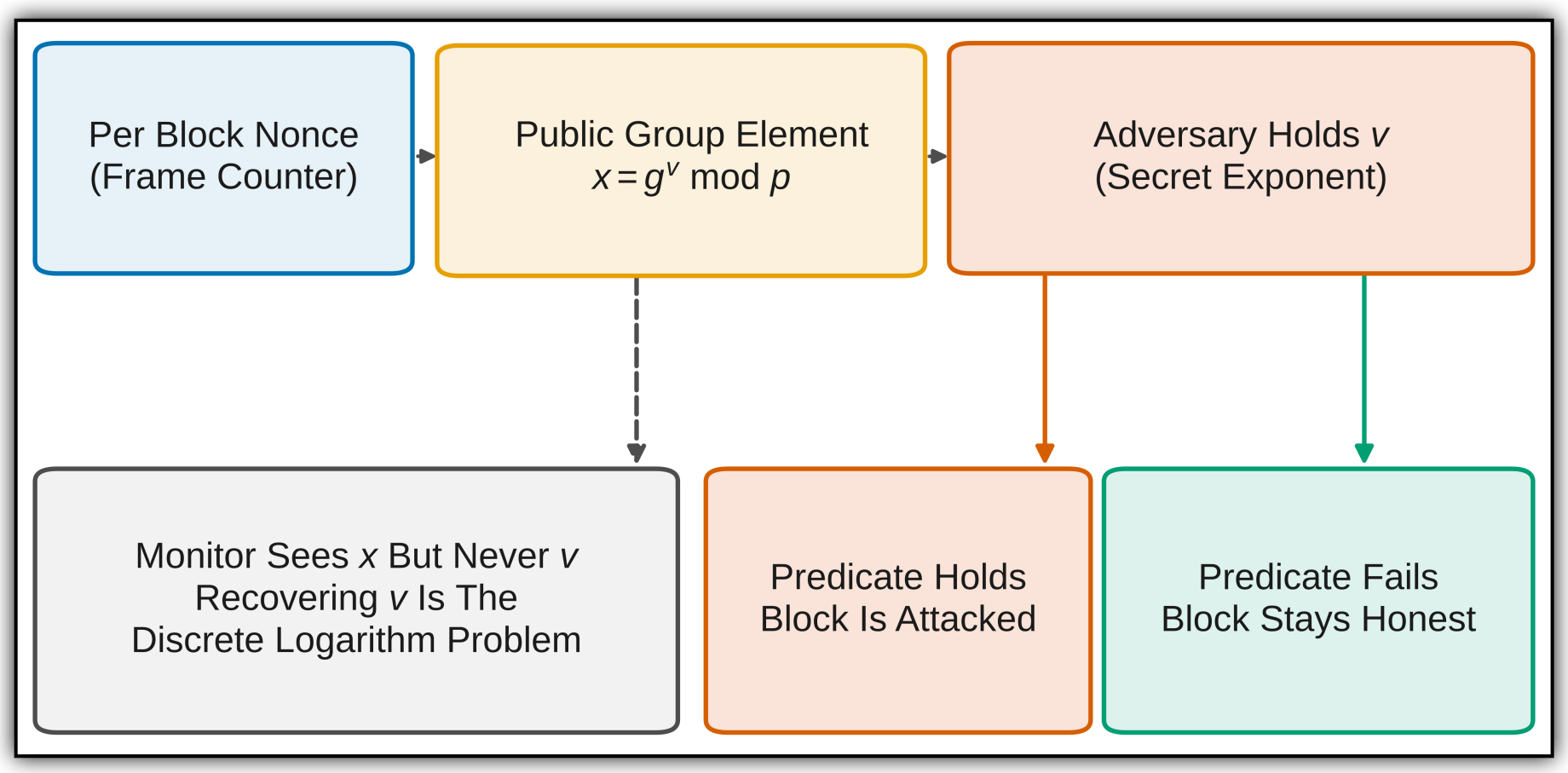

***Fig. 2*** *The camouflage gate maps a per-block nonce to $x = g^v \bmod p$, and the adversary attacks whenever its secret exponent $v$ satisfies the predicate*

The feature vector a monitor holds for a block is its telemetry together with $x$. Under a stealth primitive the telemetry of an attacked block and an honest block are drawn from the same distribution, so the telemetry carries no information about the label, and every bit of label information sits in $x$ behind the discrete logarithm. That is the whole construction, and the rest of this section establishes what it costs each side.

## 4.2 Classical undetectability

The claim to establish is that no efficient classical monitor predicts $a(x)$ from $x$ better than chance by a non-negligible margin, under the assumption that the discrete logarithm problem is hard in the chosen family of groups. Stating that claim precisely requires the usual notion of advantage, so for a classifier $\mathcal{A}$ and security parameter $n$, with the group and generator sampled from the family and $v$ uniform, I write

$$\mathrm{Adv}_{\mathcal{A}}(n) = \left| \Pr_{x} \left[ \mathcal{A}(x) = a(x) \right] - \frac{1}{2} \right|,$$

and the claim is that $\mathrm{Adv}_{\mathcal{A}}(n)$ is negligible in $n$ for every probabilistic polynomial-time $\mathcal{A}$.

**Theorem 1 (classical undetectability, conditional).** *Let the group family be one in which the discrete logarithm problem is hard, and let $\mathcal{A}$ be any probabilistic polynomial-time classifier that reads $x$ alone. Then the advantage $\mathrm{Adv}_{\mathcal{A}}(n)$ is negligible in $n$ for the interval predicate $a_{\mathrm{int}}$.*

*Proof sketch.* The function $a_{\mathrm{int}}$ is the interval-membership concept of Liu, Arunachalam and Temme evaluated on the group element $x$ [14]. A classifier with non-negligible advantage on this concept contradicts the classical-hardness half of their result, which rests in turn on the Blum and Micali theorem that the most significant bit of the discrete logarithm is as hard as the discrete logarithm itself [42]. Hardness assumptions of this shape, and the advantage notion used here, are standard in the discrete logarithm literature and are surveyed by Boneh in the setting of the decision Diffie-Hellman problem [43]. The same argument covers $a_{\mathrm{band}}$ whenever the bands are unions of intervals with endpoints fixed independently of $x$, since each interval indicator reduces to the same most-significant-bit problem. □

Two conditions carry this argument and both have to hold in the construction and cannot simply be assumed into it. The stealth footprint must leak nothing, because if attacked and honest blocks differed in any efficiently detectable telemetry statistic then a monitor would separate them directly and the hardness claim would be void. That is why the camouflaged datasets in Section 8 draw attacked-block telemetry from the honest distribution, and why I test for statistical indistinguishability instead of asserting it. The public value must also carry no shortcut, which is why the group element is presented raw and the label depends only on its discrete logarithm, leaving an efficient classical learner with the discrete logarithm problem and nothing easier. I test that directly by training a battery of classical models on the public value and its transforms and confirming they all sit at chance.

## 4.3 Quantum detectability and what the fidelity kernel can represent

A quantum kernel with the matching feature map learns $a(x)$ efficiently, and on a fault-tolerant device that feature map recovers $v = \log_g x$ from $x$ using Shor's algorithm [32], with phase estimation over the group as the subroutine that turns the period structure into a readable register [44, 45], before writing $v$ into amplitudes proportional to the characters of $\mathbb{Z}_M$,

$$\phi(x) = \frac{1}{\sqrt{K}} \left[ \cos\left( \frac{2\pi k v}{M} \right),\ \sin\left( \frac{2\pi k v}{M} \right) \right]_{k=1}^{K}, \qquad v = \log_g x,$$

normalised to unit length, with $K$ the number of frequency pairs. That letter carries two jobs in what follows, standing for the pair count when it appears alone and for the kernel when it appears with two arguments, and I hold to that convention throughout. The kernel is the state-overlap fidelity $K\left(x_i, x_j\right) = \left| \langle \phi\left(x_i\right), \phi\left(x_j\right) \rangle \right|^2$, which is exactly what an overlap circuit measures on hardware, and a support vector machine on that kernel separates the class [11, 29]. This is the quantum half of the separation of Liu, Arunachalam and Temme in the covariant-kernel form of Glick and colleagues [14, 33], specialised to the predicates above.

Two structural facts about this kernel govern the experiments, and the second of them corrects an intuition that is easy to hold and wrong.

**Proposition 1 (covariance).** *The overlap depends only on the difference of the exponents,*

$$\langle \phi(x_i), \phi(x_j) \rangle = \frac{1}{K} \sum_{k=1}^{K} \cos\left( \frac{2\pi k \left(v_i - v_j\right)}{M} \right),$$

*so the kernel is invariant under the simultaneous shift $v \mapsto v + c$ for any $c$.*

*Proof.* Expand the inner product and apply the identity $\cos A \cos B + \sin A \sin B = \cos(A - B)$ term by term, then divide by the normalisation $K$. □

**Proposition 2 (single-frequency degeneracy).** *For $K = 1$ the fidelity kernel satisfies $K(v, u) = K(v + M/2, u)$ for every group element $u$ in the training set. Consequently no classifier built on this kernel separates any predicate whose value differs between $v$ and $v + M/2$, and the interval predicate $a_{\text{int}}$ is exactly such a predicate.*

*Proof.* By Proposition 1 with $K = 1$ the fidelity is $\cos^2(2\pi(v - u)/M)$, and shifting $v$ by $M/2$ sends the argument to $2\pi(v - u)/M + \pi$, so that squaring restores the original value because $\cos(\theta + \pi) = -\cos\theta$. The two points therefore have identical kernel columns, and any kernel machine consequently assigns them the same score while their labels differ. □

The degeneracy breaks as soon as a second frequency is present, because shifting by $M/2$ multiplies the $k$-th term by $(-1)^k$, and a sum mixing even and odd $k$ is not invariant under that operation. I checked both propositions numerically on the group used in the experiments, where the $K = 1$ deviation is at the level of floating-point noise, $1.2 \times 10^{-15}$, while $K = 2$ and $K = 4$ reach deviations of order one. Figure 3 shows the effect directly, plotting the kernel against the exponent difference for three choices of $K$. I use $K = 4$ throughout, which carries the interval and band predicates comfortably, and I note that the residue predicate needs frequency pairs aligned to its divisors before the kernel can represent it at all.

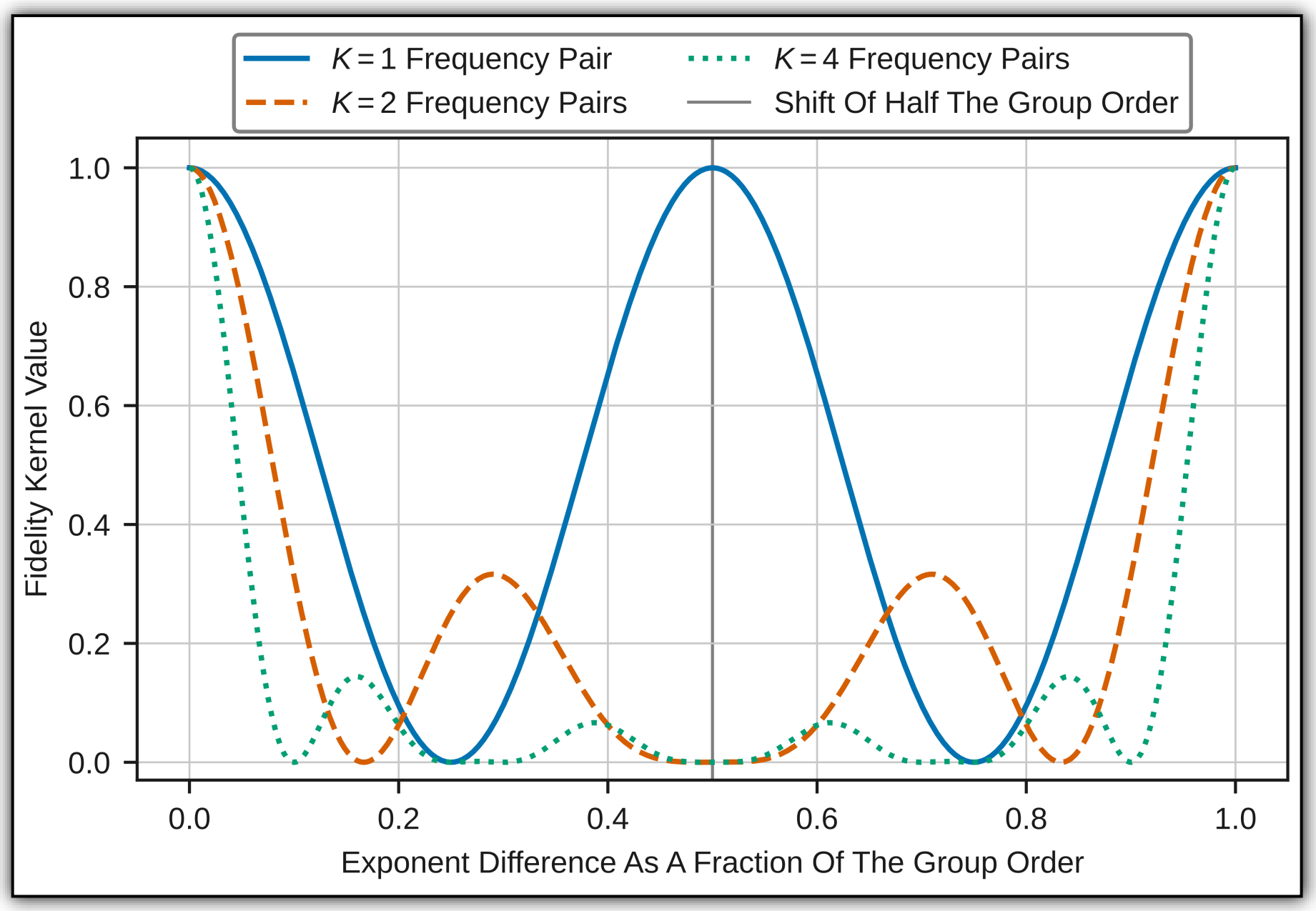

***Fig. 3*** *The fidelity kernel against exponent difference for several frequency-pair counts, showing that a single pair cannot distinguish points half a group apart*

## 4.4 What the reduction does and does not give

The separation is conditional on the hardness of the discrete logarithm problem and it is asymptotic, which together bound every claim in this paper. At a small prime the discrete logarithm falls to brute force, so the classical-hardness claim does not apply there, and the data in Section 8 shows precisely that behaviour, with a classical monitor recovering structure at 6 bits and losing it as the group grows. Following Gyurik and Dunjko, the honest form of the statement is a separation in the evaluation regime, where an efficiently computable classical hypothesis cannot compute the labels [35]. Lifting it to a fully general learning separation from textbook discrete logarithm hardness alone requires a stronger assumption than I invoke. I therefore state Theorem 1 as conditional and evaluation-regime, and I let the measurements carry the learning claim rather than overstating an unconditional result.

The choice of the discrete logarithm over a lattice problem is deliberate, and it deserves a sentence because a lattice gate would tie the construction to a post-quantum assumption and would therefore be more interesting operationally. The separation I reduce to is built on the discrete logarithm and its quantum feature map is a Shor circuit, so the reduction is available off the shelf. Learning with errors [46] is a different situation, since it is not broken by Shor, and the one sense in which it is efficiently learnable requires quantum superposition samples of the secret, an access model that simply does not exist for public classical telemetry [47]. Constructing a learning separation for a lattice gate from classical samples would be a research effort in its own right, well beyond reusing an existing theorem, so I keep the discrete logarithm gate and flag the lattice version as future work. The resource cost of the fault-tolerant feature map at cryptographic scale is the other half of this picture, and it is large: current estimates for discrete logarithms in finite fields and for comparable factoring instances run to millions of physical qubits, although that figure has fallen substantially as the constructions improved [32, 48, 49, 50].

## 5. The Quantum-Kernel Detector

The detector is the covariant fidelity kernel of Section 4.3 followed by a support vector machine. For each block it takes the public value $x$, recovers the discrete logarithm $v$, forms the character feature vector $\phi(v)$, and computes the kernel against the training set, after which the support vector machine is fitted on the same held-out split as the classical baseline so the two detectors are scored on identical data. Figure 4 shows both paths side by side, and Figure 5 shows the feature map at the circuit level, including the overlap construction that the hardware run in Section 9 actually executes. In that circuit H marks a layer of Hadamard gates and QFT the quantum Fourier transform, whose inverse reads the estimated phase out of the counting register.

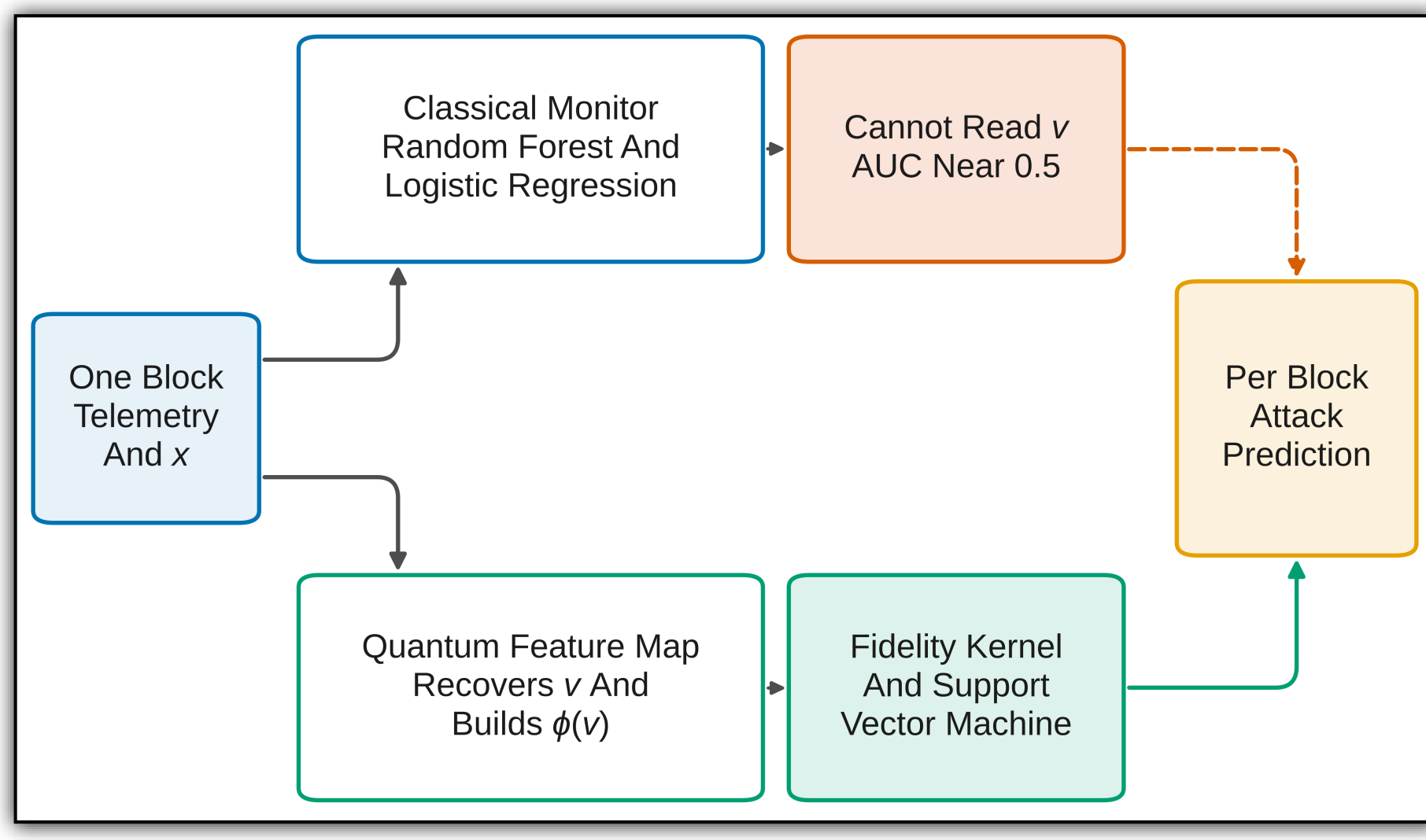

***Fig. 4*** *The two detectors compared on the same block, where only the quantum path recovers the discrete logarithm and so separates the camouflaged class*

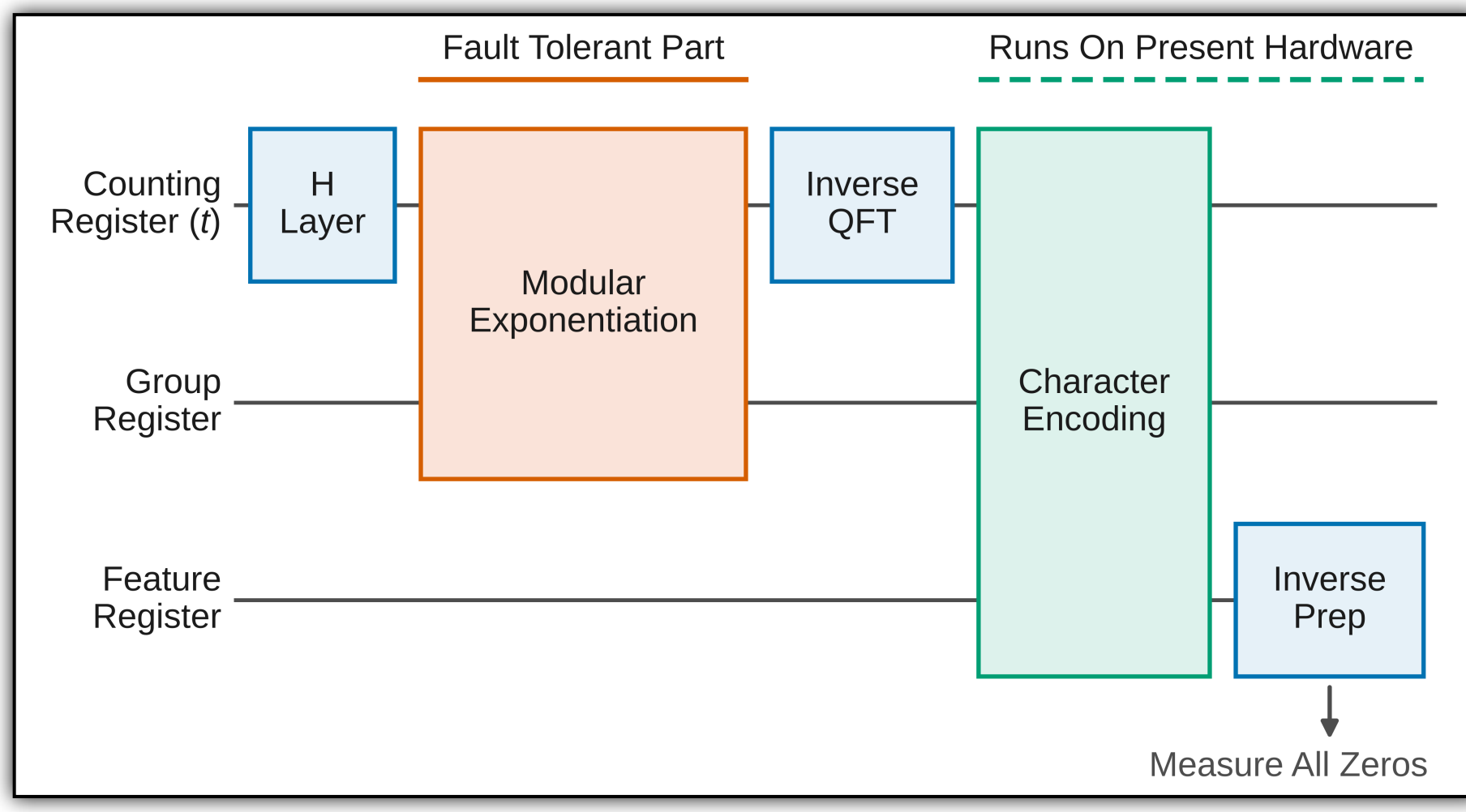

***Fig. 5*** *The feature map and overlap circuit in outline, where modular exponentiation needs fault tolerance while the encoding and overlap measurement run on present hardware*

I implement the detector as a simulation of that kernel, in two forms that the experiments switch between one variable at a time. The idealised form uses the exact exponent, obtained at these group sizes by a brute-force table, which is the exact classical emulation of the quantum discrete logarithm subroutine, and computes each fidelity in closed form. The realistic form corrupts the exponent the way a device would and estimates each kernel entry from a finite number of measurement shots rather than exactly. Both use the same feature map and the same fidelity, so any difference between them is attributable to the degradation and not to a change of model.

The realistic recovery is what makes the learning carry weight, and it is worth writing down instead of gesturing at. Phase estimation with $t$ counting qubits estimates the phase $\theta = \nu/M$ to a resolution of one part in $2^t$. Writing $s = \lfloor 2^t\theta \rfloor$ and $\delta = 2^t\theta - s$ for the integer bin and the fractional remainder, the measured bin is

$$\widehat{s} = \begin{cases} s & \text{with probability } 1-\delta, \\ (s+1) \bmod 2^t & \text{with probability } \delta, \end{cases} \qquad \widehat{\nu} = \frac{M\widehat{s}}{2^t},$$

which reproduces the property that a phase falling between two bins lands on either one with a probability set by how close it is to each [44, 45]. Gate and readout error are modelled on top of this as a depolarising replacement, so with probability $p_{\text{dep}}$ the reading is discarded and replaced by a value drawn uniformly from the group. Finally the kernel entry itself is estimated from $N_{\text{shots}}$ repetitions of the overlap circuit, so each entry carries binomial sampling noise of standard deviation $\sqrt{K(1-K)/N_{\text{shots}}}$ rather than being exact. Once the exponent is uncertain in this way the detector can no longer read a threshold off a clean number and has to generalise from corrupted features, which is the difference between a lookup table and a learning problem.

One consequence deserves to be stated head-on, since it looks at first like an unfair advantage that I have quietly granted to my own method. The recovered discrete logarithm is available to the quantum detector and withheld from the classical one, and that asymmetry is the separation itself rather than an artefact of my experimental design, because the quantum device is entitled to the discrete logarithm once Shor's algorithm computes it in polynomial time, while the classical monitor is not entitled to it because no efficient classical algorithm is known. The experiment therefore reflects an asymmetry that already exists in the complexity of the two problems, instead of manufacturing one for convenience. Figure 6 makes the consequence concrete, plotting the first character feature against the recovered exponent and showing the label falling out as a low-frequency function.

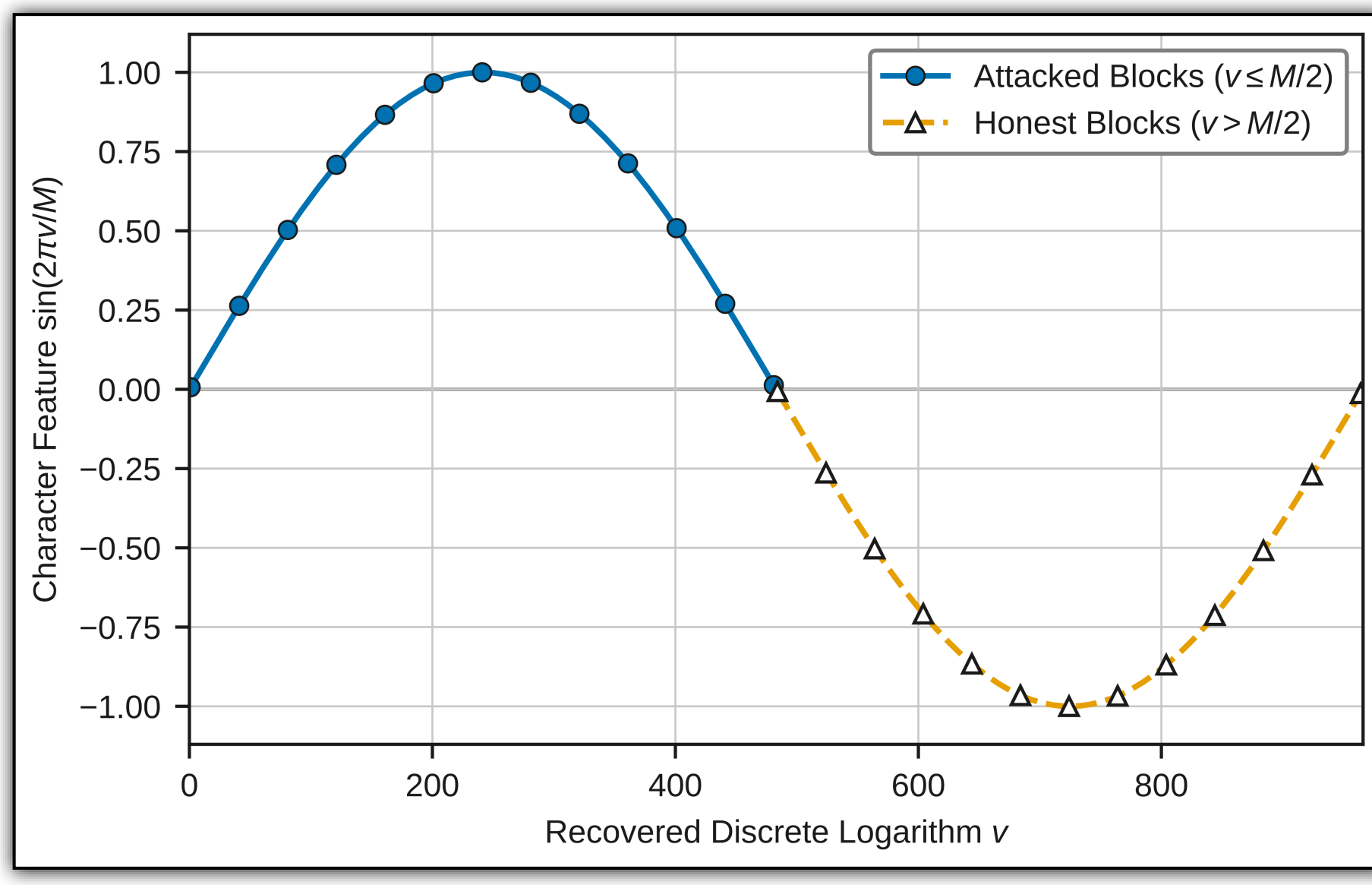

***Fig. 6*** *Once the discrete logarithm is recovered, the feature* $\sin(2\pi v/M)$ *is positive exactly on the attacked half of the group, making the label a low-frequency function*

# 6. Methodology

This section gathers the pieces of the study in one place, stating the problem, describing each component, giving the protocol in the order an operator would run it, listing the assumptions the conclusions rest on, and defining the metrics used to score everything. Collecting them here instead of scattering them through the results keeps the experimental commitments visible, so a reader who rejects one of them can see immediately which result it touches. The instantiation, the hyperparameters, and the numbers themselves follow in Sections 7 to 9.

## 6.1 Problem statement and design goals

The problem is per-block classification of a link's public telemetry under a stealth adversary that gates its attacks with a discrete logarithm predicate. Three goals shaped the design of the study and each one constrains a different part of it. The method has to reproduce the honest physics of a decoy-state BB84 link at a standard the deployed-QKD literature would accept, because a detector tested against implausible telemetry proves nothing. It has to generate an attack that is statistically indistinguishable at the block level, so the hardness result is not an artefact of a loud footprint that a better feature set would catch. And it has to place detection inside a concept class with a known quantum-classical separation, so the gap between the two detectors reflects a mechanism, with no modelling choice quietly favouring one side.

## 6.2 Components

Four components carry the study, and although each has already appeared in its own section, listing them together makes the dependencies visible. The telemetry generator feeds the attack layer, the attack layer is gated by the camouflage predicate, and the detectors see only whatever survives that chain. None of the four is a separate implementation, so a change to any one of them propagates into every result reported later.

**Telemetry generator.** A per-block Monte Carlo simulator built on the exact decoy-state model of Ma and colleagues [22], with the GLLP key rate [23] as the physics check. Detection, error, and double-click counts at each intensity are drawn as binomials over the pulses sent, so the finite-size spread of the honest envelope is produced by the sampling itself and is never assumed, with the physical parameters set to the Gobby-Yuan-Shields values [51] listed in Table 1.

**Attack layer.** The four implementation-attack primitives of Section 2.2, each with a stealth mode that holds the per-block footprint inside the honest envelope and an overt mode that leaves a visible signature. The stealth mode is credible for blinding and time-shift and partial for the other two, and Table 2 records the mapping and its honest limits.

**Camouflage gate.** The predicate families of Section 4.1, applied to the public per-block value $x = g^y \bmod p$, with a gate that is deterministic given the group and the exponent, so that labels are reproducible across seeds once the group has been fixed.

**Detectors.** A classical baseline of random forest [52] and logistic regression on the telemetry concatenated with the public value, and the covariant quantum kernel of Section 5 feeding a support vector machine [29]. Both are trained and scored on the same split, which is what makes the comparison head to head rather than suggestive.

## 6.3 Protocol

The numbered protocol below is what an operator runs to reproduce the study from end to end, and each step depends on the one before it. Step two is the one I would never skip, because a physics engine that fails the key-rate check invalidates every detector score built on top of it. The steps after that move from generation through scoring and on to the studies that stress the result rather than confirm it.

1. Fix the physical parameters (Table 1), the attack primitive, the predicate, the set of prime bit lengths, the number of blocks per dataset, and the random seed.
2. Validate the physics by reproducing the secure key rate against distance (Figure 7) and comparing the positive-rate range against the literature [22]. If the curve does not match, stop and fix the simulator before going further.
3. Generate the honest dataset from the calibrated envelope.
4. For each group size, sample a prime and a generator, then generate the overt dataset with random labels and a visible footprint and the camouflaged dataset with predicate labels and a stealth footprint.
5. Split each dataset by group element into training, validation, and test at 60/20/20, so no group element appears in two splits.
6. Fit the classical baseline and the quantum kernel on the training split and report accuracy and area under the curve on the test split.
7. Run the indistinguishability check, reporting a trained distinguisher score and per-feature Kolmogorov-Smirnov statistics on the camouflaged class.
8. For the load-bearing studies, rerun the quantum detector under finite-precision recovery with depolarising noise and shot-estimated kernels, sweep the counting qubits, and vary the labelled-block budget.
9. Repeat at a second seed, and log every run to a timestamped benchmark record fixing the parameters, the primes, and the numbers reported.

## 6.4 Assumptions

Five assumptions carry the results and each is stated here so a reader knows exactly what to break in order to break the conclusion. The physical parameters are the Gobby-Yuan-Shields values [51] as used by Ma and colleagues [22], and changing them moves the honest envelope without touching the mechanism. The stealth primitive holds the attacked-block distribution inside the honest one, which is credible for blinding and time-shift and marginal for the others. The public per-block value carries no shortcut to the label, which I test instead of assuming. The discrete logarithm problem is hard in the chosen family of groups, which is the standard assumption behind the separation I reduce to [14, 42]. And the available separation is the evaluation-regime form of Gyurik and Dunjko [35], which is what I claim and no more.

## 6.5 Metrics and reporting

The primary metrics are the area under the receiver operating characteristic curve (AUC) and the classification accuracy on held-out blocks, reported for the classical and quantum detectors side by side; the area under the curve is the appropriate primary metric here because it is insensitive to the decision threshold and to moderate class imbalance [53]. The secondary metric is the distinguisher score on the camouflaged class, which certifies that the stealth footprint gives nothing away at the block level. Every table reports both seeds, every figure that could be seed-sensitive shows both, and all runs are logged with timestamp, parameters, and raw numbers so that any single figure or table traces back to a specific execution rather than to a re-run that may have drifted.

### 6.6 Threats to validity

Three threats to internal validity are worth naming beyond the limitations collected in Section 11, of which leakage between splits is the first, and splitting by group element closes it, because at a small prime the number of distinct group elements is limited and a split by row would let a classifier memorise a lookup from $x$ to the label. Overfitting to the particular shape of the honest envelope is the second, and the two-seed protocol together with the sweep across group sizes is the guard against it. The third is the perfect quantum score itself, which I treat as a finding to be attacked instead of a result to be reported, and Sections 8.2, 8.3 and 9 are the attack.

## 7. Experimental Design

The simulator, which I call Ghillie [15], is written in Python and generates per-block telemetry by Monte Carlo, so that for each intensity the detection count in a block is drawn as a binomial over the pulses sent at that intensity, and the error and double-click counts as binomials over the detections, which makes the sampling noise that results the honest envelope itself and never an additive model layered on top. The physical parameters are the Gobby-Yuan-Shields values [51] used by Ma and colleagues [22] and are listed in Table 1. Before any attack is injected the simulator reproduces the decoy-state key rate against distance shown in Figure 7, with a positive rate out to about 142 km and a sharp cutoff, which is the check that the physics engine is faithful enough to build on.

**Table 1. Simulator physical parameters, the Gobby-Yuan-Shields values [51] as used in [22].** Source: ghillie/params.py and out_seedA/manifest.json in the Ghillie archive [15].

| Parameter | Symbol | Value |
|---|---|---|
| Fibre loss | $\alpha$ | 0.21 dB/km |
| Receiver efficiency | $\eta_{Bob}$ | 0.045 |
| Background yield | $Y_0$ | $1.7 \times 10^{-6}$ |
| Misalignment error | $e_d$ | 0.033 |
| Background error rate | $e_0$ | 0.5 |
| Signal intensity | $\mu$ | 0.48 |
| Weak decoy intensity | $\nu$ | 0.10 |
| Error-correction inefficiency | $f$ | 1.22 |
| Sifting factor | $q$ | 0.5 |
| Pulses per block | | $10^6$ |
| Blocks per dataset | | 8000 |

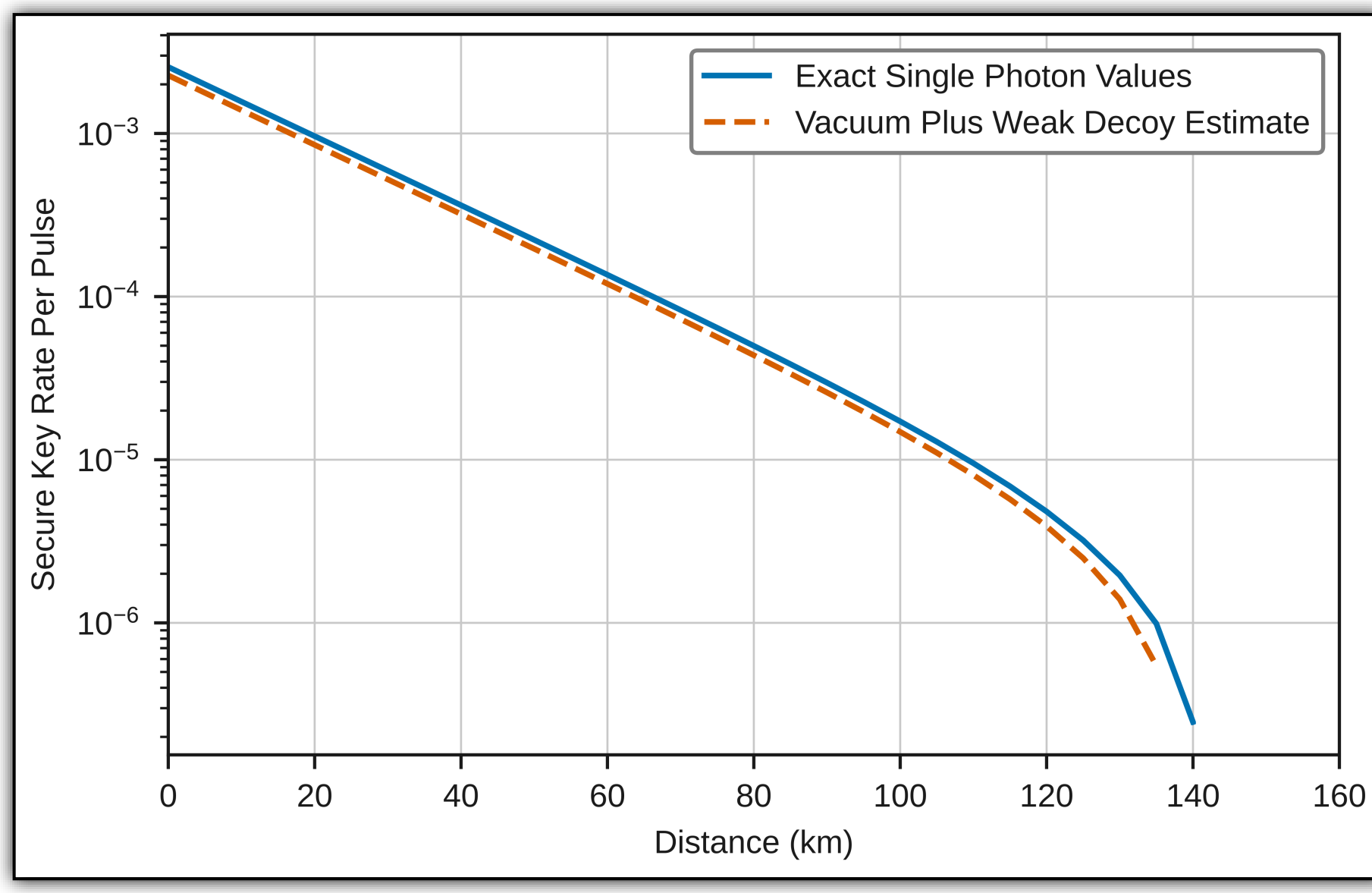

***Fig. 7*** *Secure key rate against distance at the parameters of Table 1, with a positive rate out to about 142 km that matches the decoy-state reference [22]*

Three datasets are generated at each group size, beginning with the honest dataset, which contains no attacks at all and fixes the envelope against which everything else is measured. The overt dataset attacks a random half of the blocks with a footprint large enough to leave a visible telemetry signature and serves as a positive control that any competent classical monitor should catch. The camouflaged dataset attacks according to the predicate gate with the footprint held inside the honest envelope. Camouflaged and overt datasets are generated at prime sizes of 6, 8, 10, 12, and 14 bits. Table 2 lists the attack primitives and states how faithfully each one can actually be hidden, including the two cases where the answer is only partially.

**Table 2. Attack primitives, their telemetry footprint, and how each is modelled here.** Source for the modelling column: ghillie/attacks.py in the Ghillie archive [15].

| Primitive | Telemetry footprint | Stealth fit | Modelled as |
|---|---|---|---|
| Detector blinding [4, 5] | None needed, key leaks at unchanged error rate | Strong | Default camouflage primitive |
| Time-shift [6, 7] | Timing-histogram skew, error-rate neutral | Strong | Supported |
| Photon-number-splitting [8] | Shifts signal-to-decoy gain ratio | Weak, decoy states expose it | Partial |
| Trojan-horse [9, 10] | None on the monitored channel | Outside the monitor's data | Partial |

The train, validation, and test split is by group element instead of by row, so each distinct value of $x$ is assigned to exactly one split. This matters more than it might appear, because at a small prime the number of distinct group elements is limited, and a split by row would let a classifier memorise a lookup from $x$ to the label and post a high score that says nothing whatever about hardness. Splitting by $x$ removes that path, which keeps the classical measurement honest at exactly the group sizes where the hardness claim is most fragile.

The classical baseline is a random forest with 200 trees [52] and a logistic regression with L2 penalty and 2000 maximum iterations, both from scikit-learn at default settings otherwise [54], trained on the telemetry vector concatenated with the public value. The quantum detector is the covariant fidelity kernel of Section 5 feeding a support vector machine with a precomputed kernel and $C = 1.0$ [29]. The Fourier order $K$ is 4 by default, which Proposition 2 shows to be necessary and no mere convention, and the residue predicate additionally receives the frequency pairs aligned to its divisors.

I also run an indistinguishability check between honest-label and attacked-label telemetry inside each camouflaged dataset, reported as the area under the curve of a logistic-regression distinguisher trained on telemetry alone with the public value withheld, and as the minimum Kolmogorov-Smirnov p-value across the ten telemetry features. A distinguisher near 0.5 confirms that the stealth footprint gives a trained detector nothing it can act on, which is the condition Theorem 1 depends on. I report both statistics instead of one, because a trained distinguisher can miss a marginal shift in a single feature that a per-feature test catches, and the reverse failure is equally possible. The per-feature tests are not uniformly insignificant at eight thousand blocks, and Section 8.1 reports where they are not.

Class balance in the camouflaged datasets is set by the predicate and is close to 0.5 in expectation, with observed test-split base rates recorded in the benchmark files ranging from 0.48 to 0.59 across group sizes. Chance is defined against the observed base rate for accuracy and against 0.5 for the area under the curve. Runs use 8000 blocks per dataset and $10^6$ pulses per block at two seeds, 12345 and 777, with a shorter 1500-block run at the first seed as a check that the pattern does not depend on block count. That shorter run holds the pattern, placing the classical monitor between 0.409 and 0.519 on the camouflaged class from 6 through 12 bits while the quantum kernel stays at 1.000, though its overt-attack control falls to between 0.834 and 0.893 because fewer than a fifth of the blocks leaves the classical baseline correspondingly less to learn from. The finite-precision studies reuse the same datasets and vary only the recovery precision, the shot budget, and the labelled-block count.

# 8. Results

## 8.1 The classical monitor collapses and the quantum kernel does not

The central measurement is the gap between the two detectors on the camouflaged class, plotted against group size in Figure 8 and tabulated in Table 3. In the hardness regime, meaning group sizes at which brute-force discrete logarithm is not trivial, the classical monitor reads an area under the curve near 0.5 while the quantum kernel reads 1.000, and the two seeds agree closely enough that the gap is clearly not a sampling accident. At 10, 12, and 14 bits the classical camouflaged score is 0.458, 0.516, and 0.475 at the first seed and 0.495, 0.495, and 0.496 at the second. Overt attacks are caught at every size with a classical score at or above 0.99, which is the control that separates a blind spot from a broken detector.

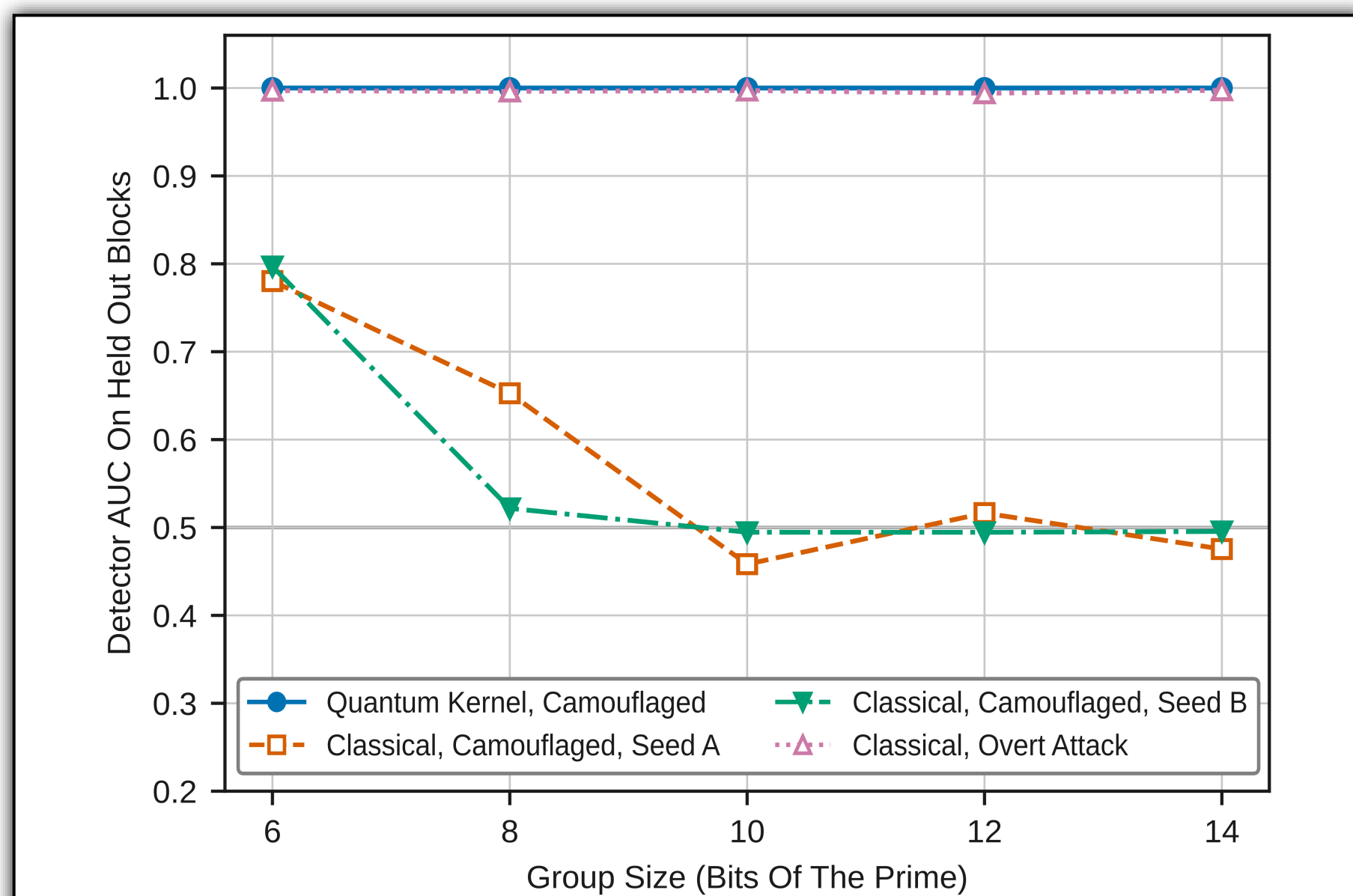

***Fig. 8*** *Detector AUC against group size, with the quantum kernel near 1.000 while the classical monitor falls to chance on camouflaged attacks yet still catches overt ones*

The smallest groups behave differently, and that difference belongs to the construction and is no defect in it. At 6 bits the classical monitor reads 0.780 and 0.797 across the two seeds, and at 8 bits it reads 0.653 and 0.522. At those sizes the discrete logarithm is small enough to brute force, so a generic learner picks up residual structure and the camouflage genuinely leaks. Blindness sets in only as the group grows past 8 bits, and an asymptotic hardness assumption predicts exactly this shape of transition, so the sweep across group size is the evidence that the transition is really there. Figure 9 shows overt and camouflaged attacks side by side at each size, and Figure 10 shows the camouflaged score converging into a band around chance across both seeds and both predicates.

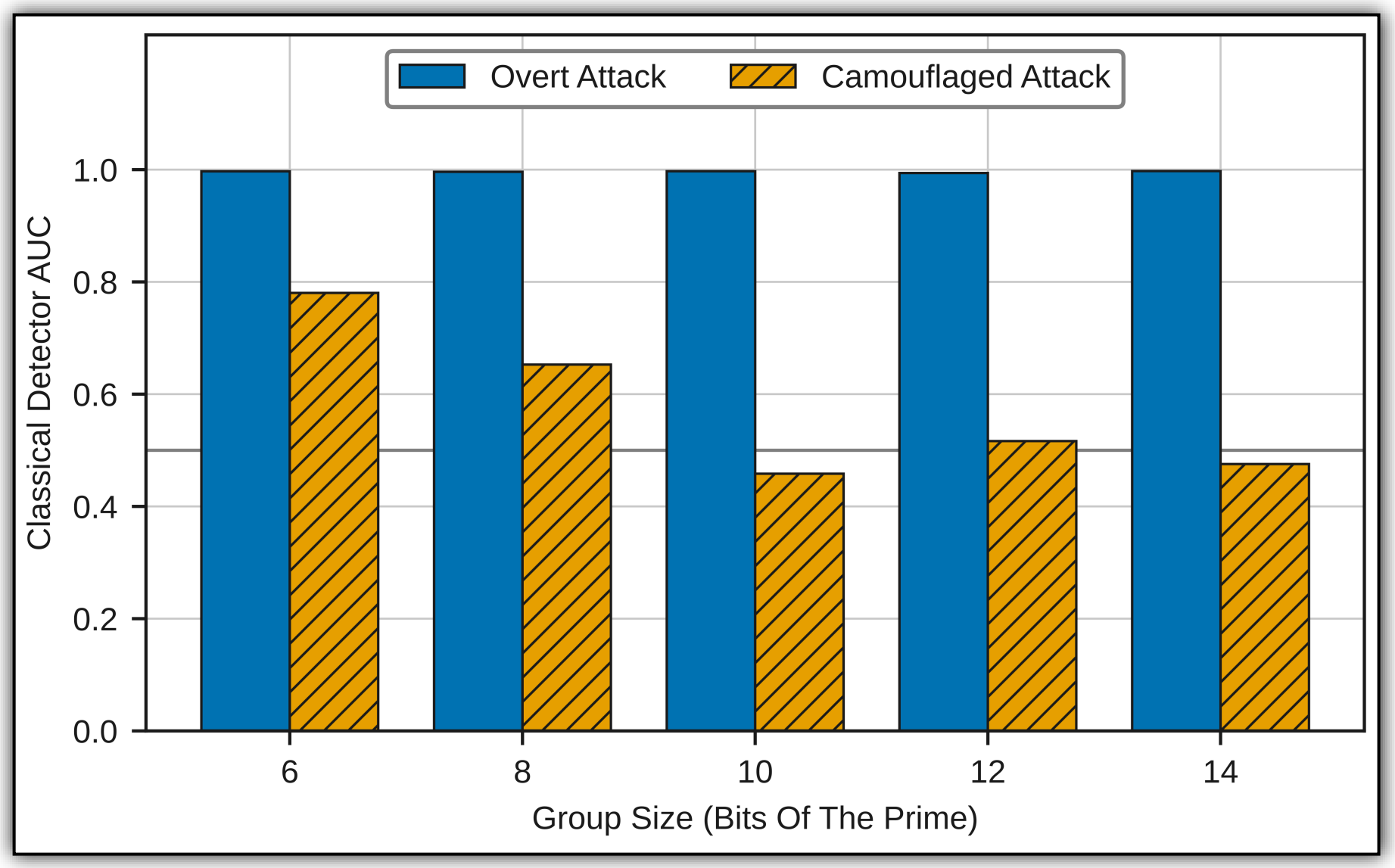

***Fig. 9*** *Classical AUC on overt and camouflaged attacks at each group size for the first seed, with camouflaged attacks missed once the group resists brute force*

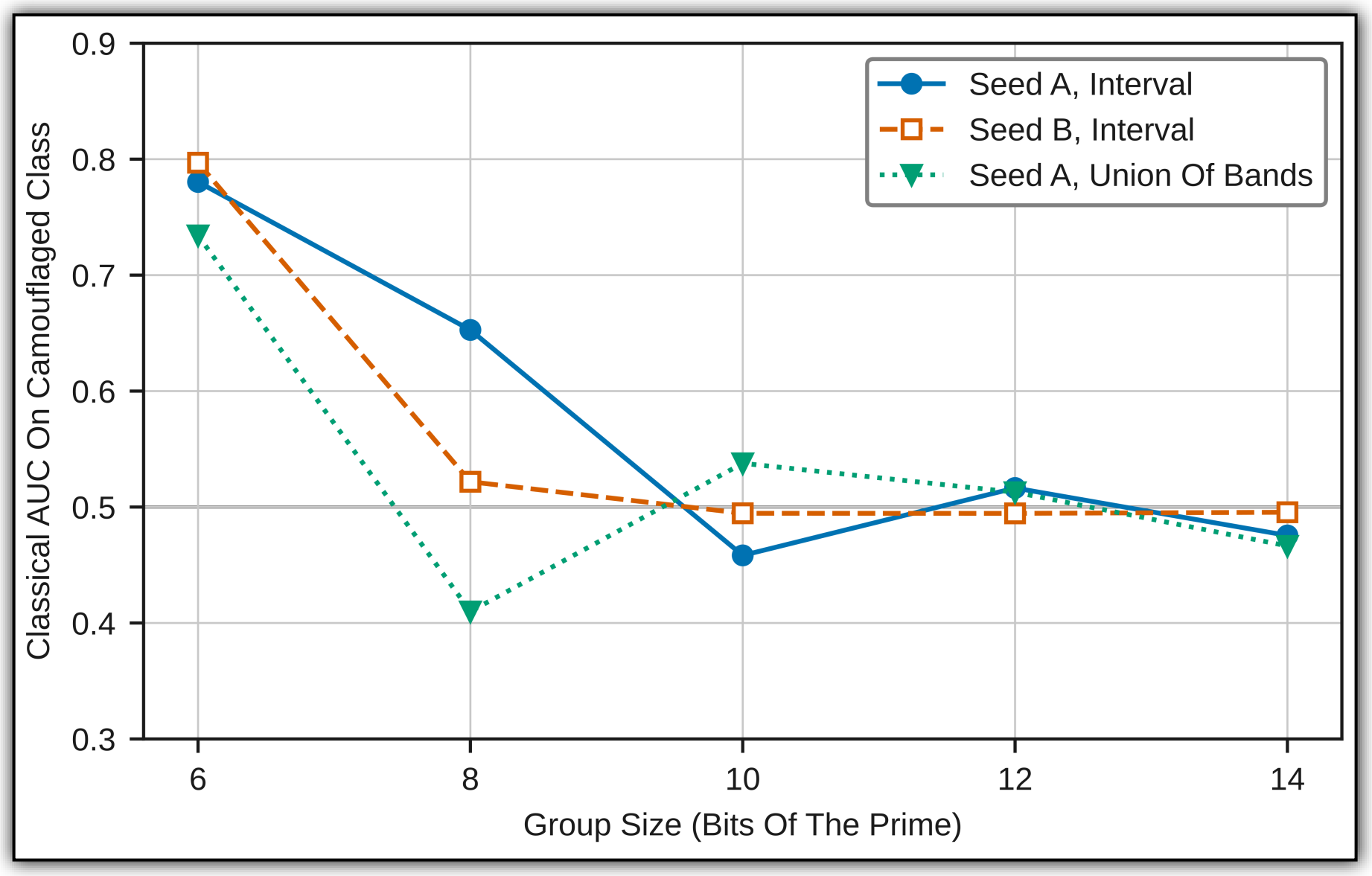

***Fig. 10*** *Classical AUC on the camouflaged class against group size across seeds and predicates, with every series falling toward the grey chance line as the group grows*

**Table 3. Detector area under the curve on held-out blocks, two seeds, 8000 blocks per dataset and** $10^6$ **pulses per block. Seed A is 12345 and seed B is 777.** Source: out_seedA/benchmarks.json and out_seedB/benchmarks.json in the Ghillie archive [15].

| Bits | Prime (A / B) | Overt, classical (A / B) | Camouflaged, classical (A / B) | Camouflaged, quantum (A / B) |
|---|---|---|---|---|
| 6 | 53 / 53 | 0.997 / 0.997 | 0.780 / 0.797 | 1.000 / 1.000 |
| 8 | 179 / 239 | 0.996 / 0.996 | 0.653 / 0.522 | 1.000 / 1.000 |
| 10 | 967 / 631 | 0.997 / 0.998 | 0.458 / 0.495 | 1.000 / 1.000 |
| 12 | 3089 / 4007 | 0.994 / 0.997 | 0.516 / 0.495 | 1.000 / 1.000 |
| 14 | 11779 / 10151 | 0.997 / 0.996 | 0.475 / 0.496 | 1.000 / 1.000 |

Accuracy follows the same pattern and is reported in Table 4, with the quantum detector at or just below 1.0 and the classical detector at chance against its own base rate in the hardness regime. The two metrics agree here because the camouflaged datasets sit close to balanced, so no threshold choice rescues the classical side. Reporting both is a guard against the objection that one summary statistic might be concealing a skew in the other.

**Table 4. Test-set accuracy on the camouflaged class, two seeds.** Source: out_seedA/benchmarks.json and out_seedB/benchmarks.json in the Ghillie archive [15].

| Bits | Classical accuracy (A / B) | Quantum accuracy (A / B) |
|---|---|---|
| 6 | 0.708 / 0.694 | 1.000 / 1.000 |
| 8 | 0.616 / 0.519 | 1.000 / 1.000 |
| 10 | 0.469 / 0.497 | 1.000 / 0.996 |
| 12 | 0.510 / 0.488 | 1.000 / 1.000 |
| 14 | 0.483 / 0.491 | 0.999 / 1.000 |

The indistinguishability of the telemetry is what licenses Theorem 1, and it is confirmed in Table 5 and Figure 11, with a single channel shown in Figure 12. A distinguisher trained to tell attacked blocks from honest ones on telemetry alone sits near 0.5 at every group size, so there is no per-block signature to find and the only handle on the label is the discrete logarithm structure of the public value. That is the empirical content of the stealth assumption, and without it the hardness argument of Section 4.2 would have nothing underneath it. The per-feature picture is less uniform than the distinguisher alone suggests, and it is worth stating exactly. Across ten telemetry features, five group sizes and two seeds, the smallest Kolmogorov-Smirnov p-value is $1.9 \times 10^{-4}$, on the decoy gain at 12 bits at the second seed, where the test statistic is 0.048 and the standardised mean difference sits below 0.10. What a test of that power resolves at eight thousand blocks is a residual difference far too small for any detector to convert into an advantage, which is what the distinguisher score of 0.522 on the same block set reflects. I treat the distinguisher as the operationally meaningful statistic while recording that a sufficiently sensitive per-feature test does eventually reach the floor of the simulator's stealth mode.

**Table 5. Statistical indistinguishability of attacked and honest telemetry on the camouflaged class. A distinguisher near 0.5 means the two are not separable at the block level.** Source: out_seedA/benchmarks.json and out_seedB/benchmarks.json in the Ghillie archive [15].

| Bits | Distinguisher AUC (A / B) |
|---|---|
| 6 | 0.478 / 0.506 |
| 8 | 0.500 / 0.505 |
| 10 | 0.502 / 0.492 |
| 12 | 0.514 / 0.522 |
| 14 | 0.512 / 0.494 |

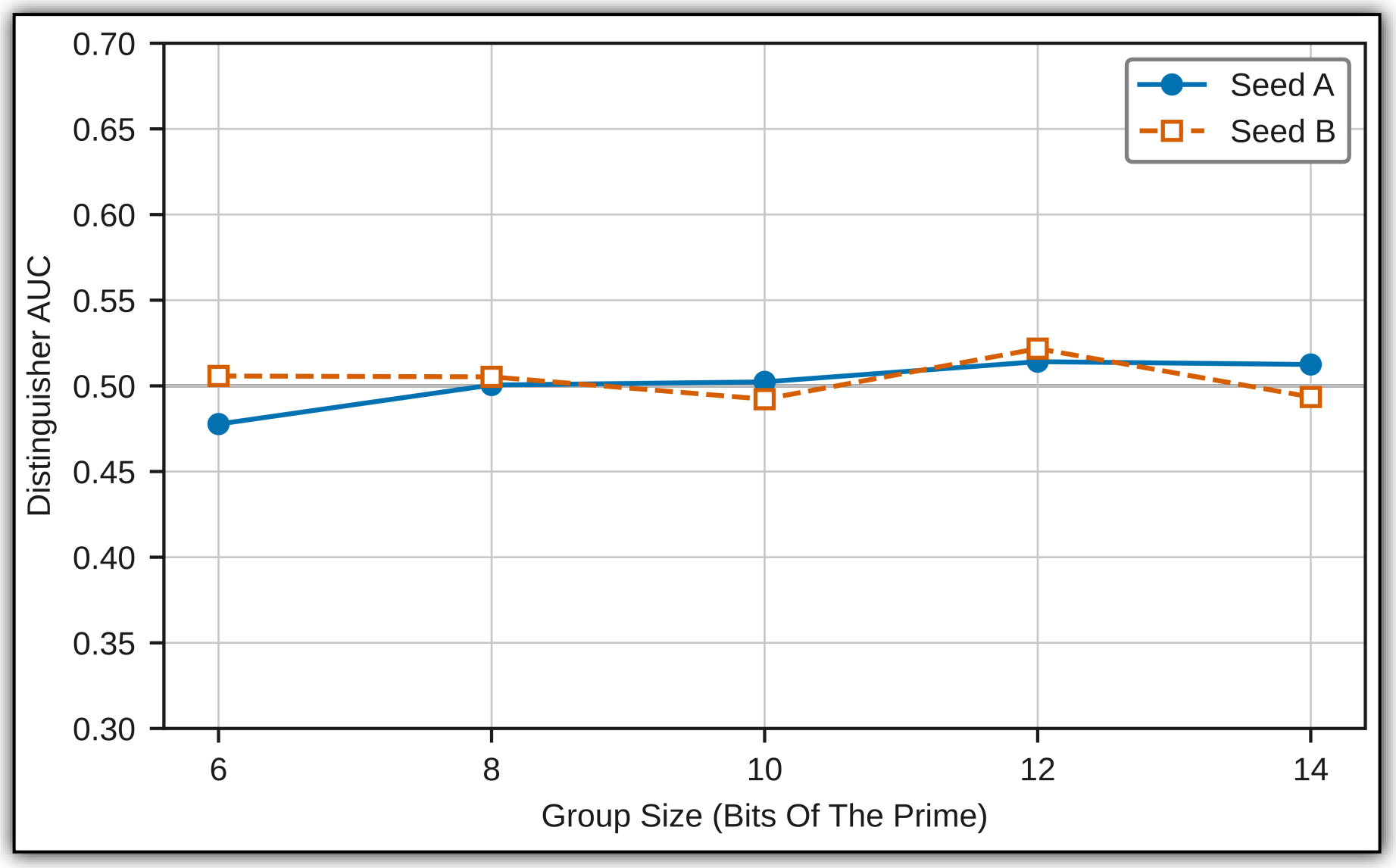

***Fig. 11*** *Distinguisher AUC for separating attacked from honest telemetry, which stays near chance at every group size on both seeds*

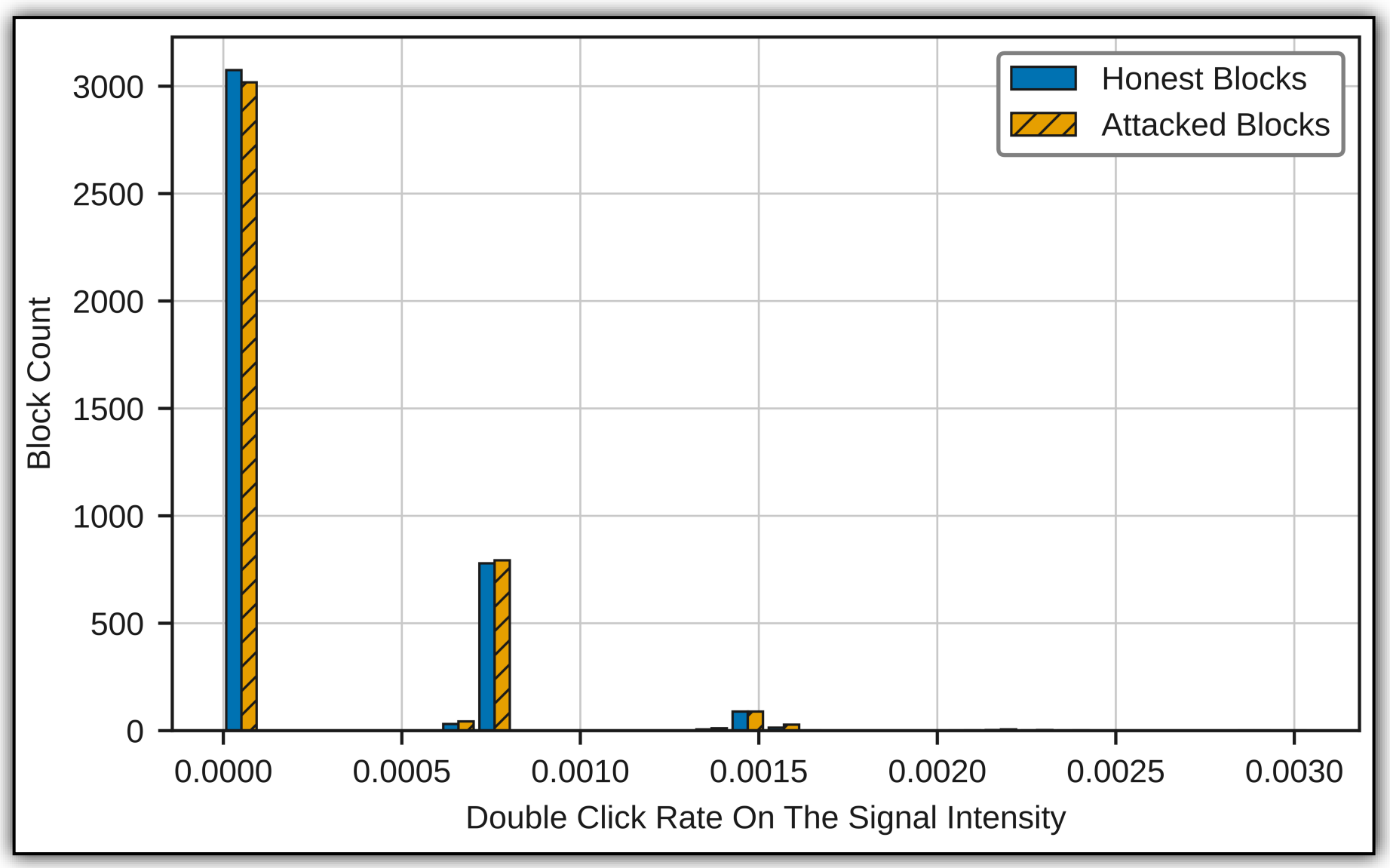

***Fig. 12*** *Honest and attacked telemetry have nearly identical distributions on the double-click channel at a ten-bit group, leaving a telemetry-only monitor nothing to separate*

The quantum score of exactly 1.000 in Table 3 needs an explanation instead of a victory lap, and the explanation is mundane. Given an exact exponent, the covariant feature map encodes the interval label in a low-frequency subspace, and on clean labels the support vector machine separates that subspace without error. The honest content of the comparison is the contrast with a classical monitor holding the same data minus the discrete logarithm. Whether the quantum side is doing anything beyond inheriting a free answer from an exact exponent is a fair question, and the rest of this section is my attempt to answer it against my own interest.

## 8.2 Making the learning load-bearing

A perfect score at an exact exponent is easy to dismiss, and the dismissal is reasonable on its face. If Shor hands the detector the exact value of $v$ and the label is a single threshold on $v$, then one comparison settles the matter and nothing deserving the name learning has occurred. I take that objection apart by changing one variable at a time and reporting what survives. Table 6 gives the ablation at 12 bits, in which the idealised row is the generous case from Table 3, using an exact exponent together with the interval predicate. The realistic row keeps the interval predicate but recovers the exponent through the model of Section 5, with four counting qubits, a two percent depolarising replacement, and each kernel entry estimated from 2048 shots. The hardened row keeps the exact exponent while swapping in the union-of-bands predicate, and the last row combines both degradations, which makes it the closest of the four to a deployment setting.

**Table 6. Ablation of the quantum detector at 12 bits, two seeds. The classical monitor is shown for reference and stays at chance in every row. Only the named variable changes between rows.** Source: benchmarks/ablation_20260914_015419.csv (seed A) and benchmarks/ablation_20260914_015813.csv (seed B) in the Ghillie archive [15].

| Mode | Predicate | Recovery | Classical AUC (A / B) | Quantum AUC (A / B) |
|---|---|---|---|---|
| Idealised | Interval | Exact exponent, exact kernel | 0.490 / 0.527 | 1.000 / 1.000 |
| Realistic | Interval | 4 qubits, depolarising, 2048 shots | 0.490 / 0.527 | 0.978 / 0.983 |
| Hardened | Union of bands | Exact exponent, exact kernel | 0.478 / 0.479 | 1.000 / 1.000 |
| Hardened and noisy | Union of bands | 4 qubits, depolarising, 2048 shots | 0.478 / 0.479 | 0.916 / 0.927 |

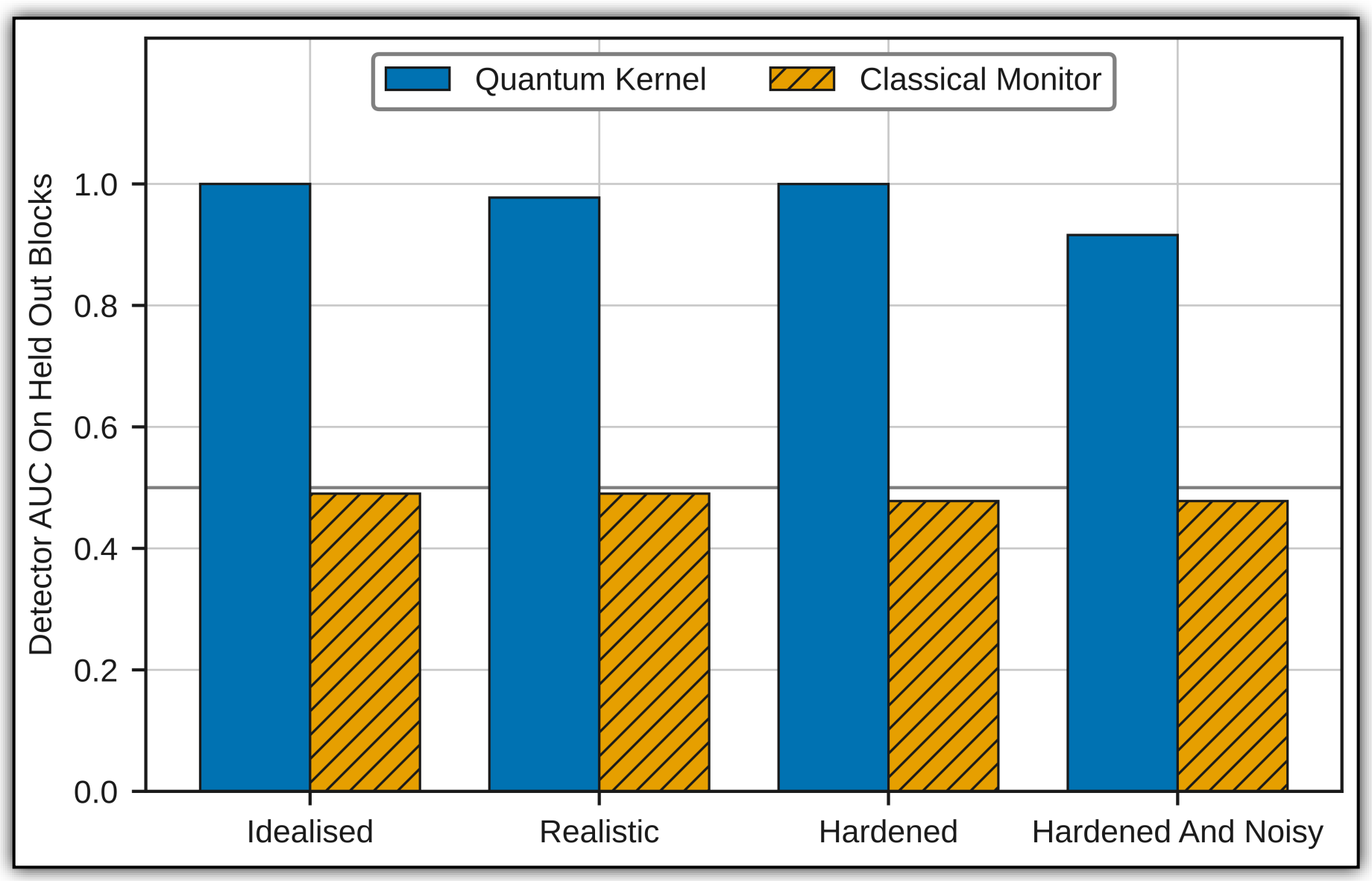

***Fig. 13*** *The ablation of Table 6 at the first seed, where the quantum score falls as recovery becomes realistic and the predicate hardens while the classical monitor stays at chance*

The pattern in Figure 13 is the one an honest account predicts, in that the perfect score is real but fragile, and it degrades as soon as the exponent stops being exact, to about 0.98 under a realistic interval recovery and to about 0.92 when the predicate is hardened as well. The union predicate degrades faster under noise than the interval does, which follows from the representation argument in Section 4.3, since a boundary needing several frequencies to describe is more exposed to a corrupted feature than one resting on a single low frequency. Through all four rows the classical monitor does not move off chance, so the gap between the detectors is not an artefact of the generous setup. I found the size of the hardened and noisy drop reassuring rather than disappointing, because a score that stayed pinned at 1.000 under noise would have been the more suspicious outcome and would have suggested I had left a shortcut in the pipeline.

The precision sweep in Figure 14 traces the realistic recovery across the number of counting qubits with everything else held fixed. The quantum score rises smoothly from 0.862 and 0.844 at two counting qubits to 0.990 and 0.988 by five, on the two seeds respectively, and it flattens out above five where added precision buys almost nothing. The classical monitor stays pinned at 0.458 throughout, because the exponent is unavailable to it at any precision whatsoever. This is the clearest single picture of where the learning does its work, because at low precision the recovered exponent is wrong often enough that the detector must generalise from corrupted features, and the score then measures how much signal survives that corruption, while as precision grows the score climbs steadily toward the clean-label ceiling.

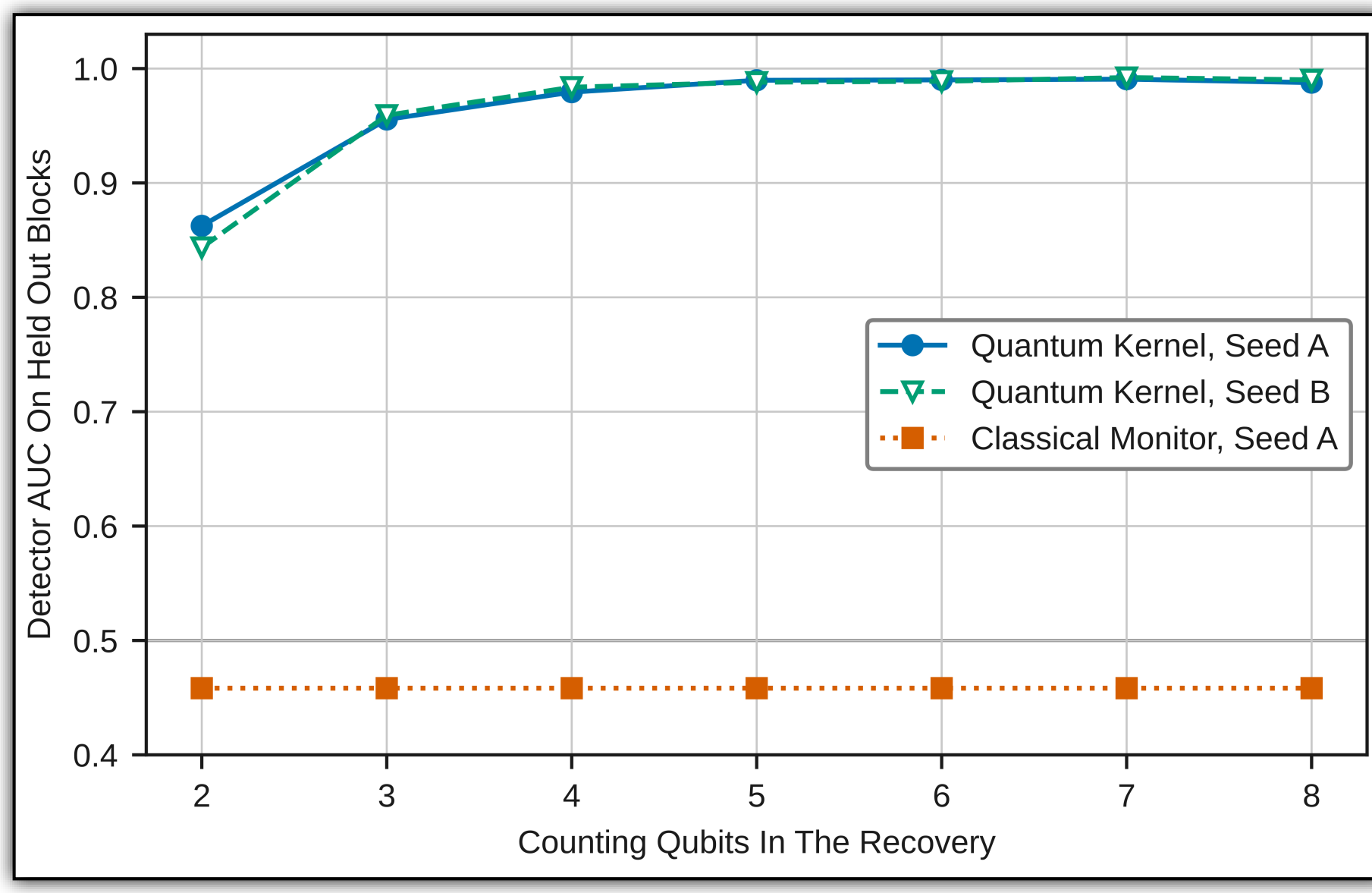

***Fig. 14*** *Detector AUC against counting qubits, with the quantum score rising from about 0.85 to 0.99 as precision grows while the classical monitor stays at chance*

**Table 7. Detector AUC against the number of counting qubits used in the recovery, at a ten-bit group with the interval predicate and a two percent depolarising replacement. The classical monitor never sees the exponent, so its score does not move with precision.** Source: benchmarks/sweep_20260914_015630.csv (seed A) and benchmarks/sweep_20260914_020033.csv (seed B) in the Ghillie archive [15].

| Counting qubits | Quantum, A | Quantum, B | Classical, A |
|---|---|---|---|
| 2 | 0.862 | 0.844 | 0.458 |
| 3 | 0.956 | 0.959 | 0.458 |
| 4 | 0.979 | 0.984 | 0.458 |
| 5 | 0.990 | 0.988 | 0.458 |
| 6 | 0.990 | 0.989 | 0.458 |
| 7 | 0.991 | 0.992 | 0.458 |
| 8 | 0.988 | 0.990 | 0.458 |

The sample-complexity study in Figure 15 asks how much labelled telemetry each detector needs, which is the question an operator would actually ask first. Under a realistic four-qubit recovery the quantum kernel reaches 0.931 and 0.971 from twenty-five labelled blocks, settles near 0.97 by one hundred, and moves very little out to eight hundred, while the classical baseline stays between 0.477 and 0.511 at every budget I tried. The curve is close to flat, and that flatness is itself the finding and carries no defect of the experiment: at this recovery precision a small training set already pins the boundary, so a defender who can label even a modest run of blocks has what the quantum detector needs, and no amount of additional data rescues the classical side because the signal it would need is behind the discrete logarithm.

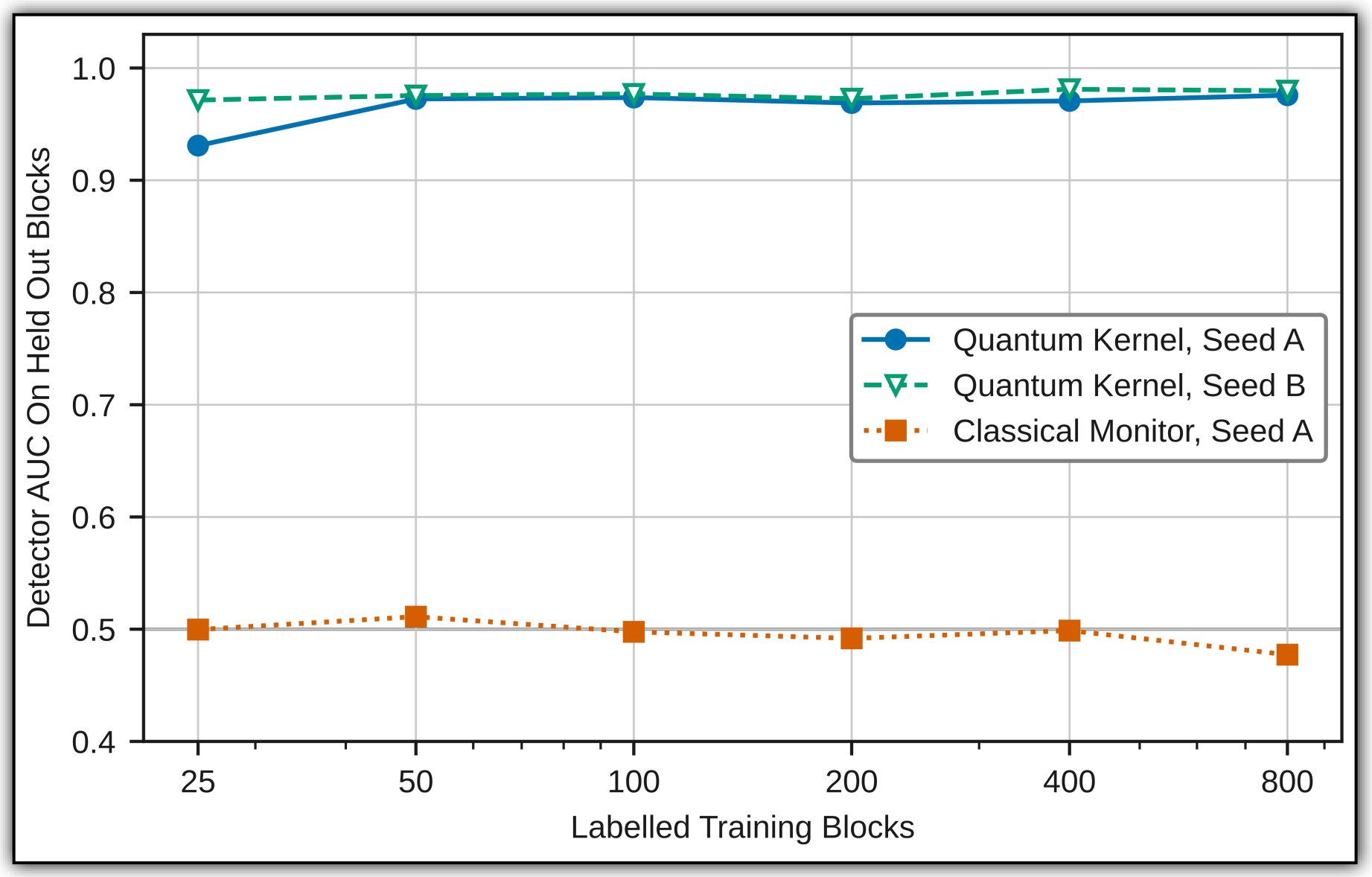

***Fig. 15*** *Detector AUC against labelled training blocks, where the quantum kernel reaches high accuracy from a small budget and the classical monitor never leaves chance*

**Table 8. Detector AUC against the number of labelled training blocks, at a ten-bit group under a realistic four-qubit recovery with 2048 shots per kernel entry.** Source: benchmarks/samples_20260914_015723.csv (seed A) and benchmarks/samples_20260914_020053.csv (seed B) in the Ghillie archive [15].

| Labelled blocks | Quantum, A | Quantum, B | Classical, A |
|---|---|---|---|
| 25 | 0.931 | 0.971 | 0.500 |
| 50 | 0.972 | 0.976 | 0.511 |
| 100 | 0.974 | 0.977 | 0.498 |
| 200 | 0.969 | 0.973 | 0.492 |
| 400 | 0.971 | 0.981 | 0.499 |
| 800 | 0.976 | 0.980 | 0.477 |

Taken together these three studies answer the objection that the perfect score invites, because the quantum advantage is not an artefact of an exact exponent paired with a trivial threshold, surviving as it does finite precision, depolarising noise, a hardened predicate, and a scarce labelled budget, degrading in each case by an amount that can be read directly off the figures, while the classical monitor stays at chance throughout. What the exact-exponent case supplies is a clean ceiling, and what these three studies supply is the shape of the descent away from it, which is where a real deployment would live instead of at the top. I would rather report that descent plainly than defend a ceiling no device is going to reach.

### 8.3 The feature map has to match the predicate

The residue predicate produced the result I did not expect, and on reflection it is the most useful thing in the paper for anyone who wants to build such a detector. To isolate the question I fixed the exponent as exact, so that nothing but the representational power of the feature map could be responsible for the outcome, and then crossed the three predicate families against three feature maps at a ten-bit group on both seeds. Table 9 and Figure 16 give the result, which is stark enough that it changes how I would describe the method.

**Table 9. Quantum detector AUC for each adversary predicate against each feature map, using an exact exponent and 1200 exponents split by group element. Seed A runs at the prime 967 with generator 5 and seed B at the prime 631 with generator 3, both taken from the ten-bit datasets used elsewhere in the paper. The classical reference never leaves the 0.456 to 0.582 band in any cell and is omitted for clarity.** Source: benchmarks/alignment_20260920_043220.csv (seed A) and benchmarks/alignment_20260920_043225.csv (seed B) in the Ghillie archive [15].

| Adversary predicate | One frequency pair (A / B) | Four frequency pairs (A / B) | Four pairs plus divisor frequencies (A / B) |
|---|---|---|---|
| Interval | 0.449 / 0.451 | 1.000 / 1.000 | 1.000 / 1.000 |
| Union of bands | 0.493 / 0.467 | 0.998 / 0.998 | 0.998 / 0.998 |
| Residue pattern | 0.523 / 0.461 | 0.545 / 0.442 | 0.982 / 1.000 |

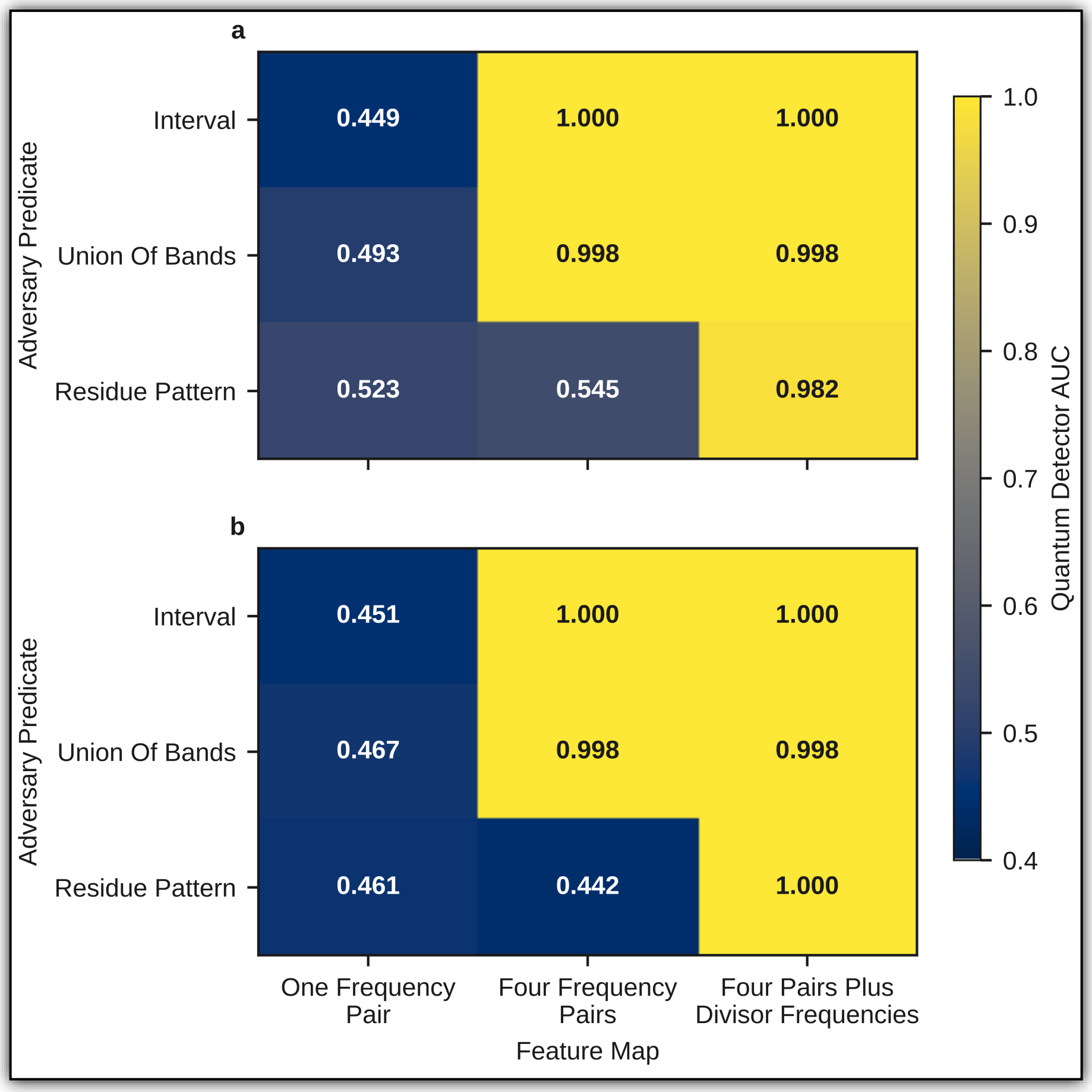

***Fig. 16*** *Quantum detector AUC for each predicate and feature map at seed A (a) and seed B (b), with the residue pattern detected only once divisor frequencies are added*

Three readings come out of that table and each one matters for a different reason. The first column is Proposition 2 measured rather than proved, since a single frequency pair leaves every predicate at chance, and the interval case sits at 0.449 exactly as the degeneracy argument predicts. The second column shows that four low frequencies carry the interval and the union comfortably on both seeds, which is why the headline results of Section 8.1 look as clean as they do. The third column is the one that should give a defender pause, because the residue predicate stays at 0.545 and 0.442 under a four-frequency map, which is no better than the classical monitor on the same data, and only reaches 0.982 and 1.000 once the map is handed the frequencies tied to the small primes the adversary chose.

What this means in practice is that the quantum advantage is conditional on knowledge the defender may not possess. A covariant kernel earns its separation only when its feature map spans the structure the adversary selected, and a mismatched map inherits nothing at all, which is the practical form of the alignment condition that Kübler and colleagues derive from the kernel spectrum [31]. I think this is the honest limit of the whole construction, and I would rather state it as a headline finding than bury it in a limitations paragraph, because a defender who guesses the predicate family wrongly gets a detector that performs no better than the classical monitor it was meant to replace.

## 9. Hardware Run on IBM Quantum Processors

The kernel entry the detector needs is a state-overlap fidelity, and that fidelity is what an overlap circuit measures directly. Preparing $\phi(x_i)$, applying the inverse of the preparation of $\phi\left(x_j\right)$, and reading the probability of the all-zeros outcome returns the fidelity in the infinite-shot limit, and an estimate from finitely many shots is the sampled kernel used in Section 8.2. I ran that circuit on IBM hardware to establish that it behaves on a real device. It is a feasibility probe and nothing more, because the asymptotic separation needs fault tolerance and sits far out of reach at any group size where the discrete logarithm is genuinely hard [55].

The setup is deliberately small so that it fits inside the free-tier runtime budget, and I fixed a ten-bit prime, $p = 571$ with generator 3, then chose ten exponents spanning both classes of the predicate, which gives a ten-by-ten kernel and one hundred overlap circuits per run. The interval and union feature maps use four frequency pairs and encode into three qubits after padding to a power of two, while the residue map needs its divisor-aligned frequencies and encodes into four. Each circuit was transpiled to the backend's native gate set and coupling map with the runtime preset pass manager at optimisation level one, then submitted through the SamplerV2 primitive in job mode, which is what the free Open Plan supports. Runs were executed on two 156-qubit Heron r2 processors, ibm_kingston and ibm_fez, across the three predicates and three shot budgets, using Qiskit 2.5.2 with qiskit-ibm-runtime 0.49.0. Every run records the backend name, the job identifier, and a calibration snapshot taken at execution time, so each number below traces to a specific execution on a specific device.

Figure 17 is the core check, plotting each measured kernel entry from the interval runs on both processors against the value it should take. The points track the diagonal across the full range, with a mild downward pull at the top which is the expected signature of readout and gate error depressing a true fidelity of one toward roughly 0.85. Table 10 quantifies the agreement across all six runs, where the mean absolute difference between measured and exact entries sits between 0.016 and 0.034, and a support vector machine trained on the hardware-measured kernel separates the two classes at training accuracy 1.000 in every run, including the deeper four-qubit residue circuit.

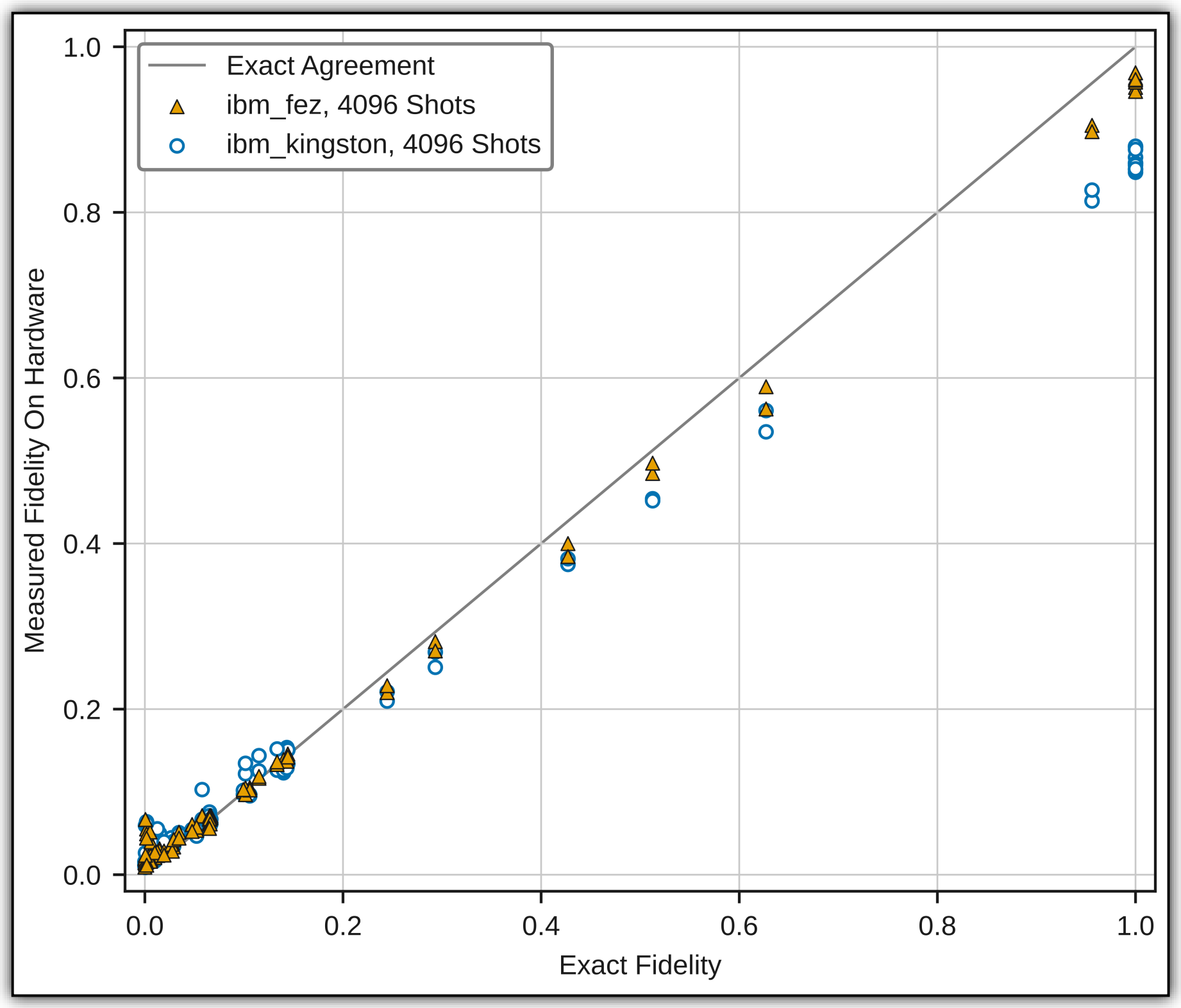

***Fig. 17*** *Measured against exact kernel entries for the interval runs on both processors, where the grey line marks exact agreement*

**Table 10. Hardware runs on IBM Heron r2 processors. Every run is a ten-by-ten fidelity kernel over one hundred overlap circuits on the prime 571 with generator 3. Agreement is the mean absolute difference between measured and exact kernel entries, and the final column is the training accuracy of a support vector machine on the measured kernel.** Source: run_meta.json and kernel_entries.csv in the six ibm_runs folders in the Ghillie archive [15].

| Backend | Job ID | Predicate | Qubits | Shots | Mean abs error | SVM accuracy |
|---|---|---|---|---|---|---|
| ibm_fez | dajh86ni3e6s738q9f70 | Interval | 3 | 4096 | 0.018 | 1.000 |
| ibm_kingston | dajhakr9k43c73ah7n50 | Interval | 3 | 4096 | 0.034 | 1.000 |
| ibm_kingston | dajka2ni3e6s738qclbg | Union of bands | 3 | 4096 | 0.016 | 1.000 |
| ibm_kingston | dajkbavi3e6s738qcml0 | Residue | 4 | 4096 | 0.023 | 1.000 |
| ibm_kingston | dajkfghhvn6c73cuhcn0 | Interval | 3 | 1024 | 0.017 | 1.000 |
| ibm_kingston | dajkv38mhr3c73e93ch0 | Interval | 3 | 8192 | 0.023 | 1.000 |

The calibration snapshots recorded at submission place these runs in a normal operating window for both devices, and Table 11 lists them beside the physical qubits that the transpiler actually selected, which I read from the circuit diagrams exported by the IBM Quantum platform for each job. Something in that table cuts against the explanation I first reached for, because the run with the best median readout error, the first interval run on ibm_kingston at 0.68 percent, produced the worst agreement of all six at 0.034, while ibm_fez at 0.92 percent readout error produced one of the best. Device medians therefore fail to predict which run tracks the exact kernel most closely, and the qubit column supplies a better candidate explanation.

**Table 11. Calibration snapshots captured at submission time for each hardware run, the physical qubits chosen by the transpiler, and the agreement achieved. Coherence times and readout error are device medians, taken across all 156 qubits, while the qubit column names the three or four actually used.** Source: the calibration snapshots in the six ibm_runs folders and the transpiled circuit exports in ibm_workloads, all in the Ghillie archive [15].

| Backend | Predicate | Shots | Qubits used | T1 (μs) | T2 (μs) | Readout error | Mean abs error |
|---|---|---|---|---|---|---|---|
| ibm_fez | Interval | 4096 | 21, 22, 23 | 136.1 | 117.8 | 0.0092 | 0.018 |
| ibm_kingston | Interval | 4096 | 140, 141, 142 | 255.7 | 125.5 | 0.0068 | 0.034 |
| ibm_kingston | Union of bands | 4096 | 20, 21, 36 | 255.7 | 125.5 | 0.0096 | 0.016 |
| ibm_kingston | Residue | 4096 | 104, 105, 106, 117 | 255.7 | 125.5 | 0.0096 | 0.023 |
| ibm_kingston | Interval | 1024 | 105, 106, 117 | 255.7 | 125.5 | 0.0096 | 0.017 |
| ibm_kingston | Interval | 8192 | 105, 106, 117 | 255.7 | 125.5 | 0.0096 | 0.023 |

The pattern in that column is hard to dismiss as coincidence, because the single worst run is also the only ibm_kingston job that the transpiler routed onto qubits 140 through 142, while every other job on that device landed either in the 104 to 117 neighbourhood or on the low-numbered qubits around 20, and all of those tracked the exact kernel within 0.023. This points at the two-qubit gate quality on whichever handful of qubits the layout pass happens to pick, rather than at any figure averaged over the whole device, and it suggests that a deployment wanting reproducible kernel estimates would pin the layout explicitly instead of letting the transpiler choose afresh on every submission. The residue circuit also confirms its own description, since the exported diagram carries a four-bit measurement register where every other run carries three, matching the extra qubit that the divisor-aligned feature map requires. Figure 18 shows the same variation directly, with every run plotted on one axis for comparison.

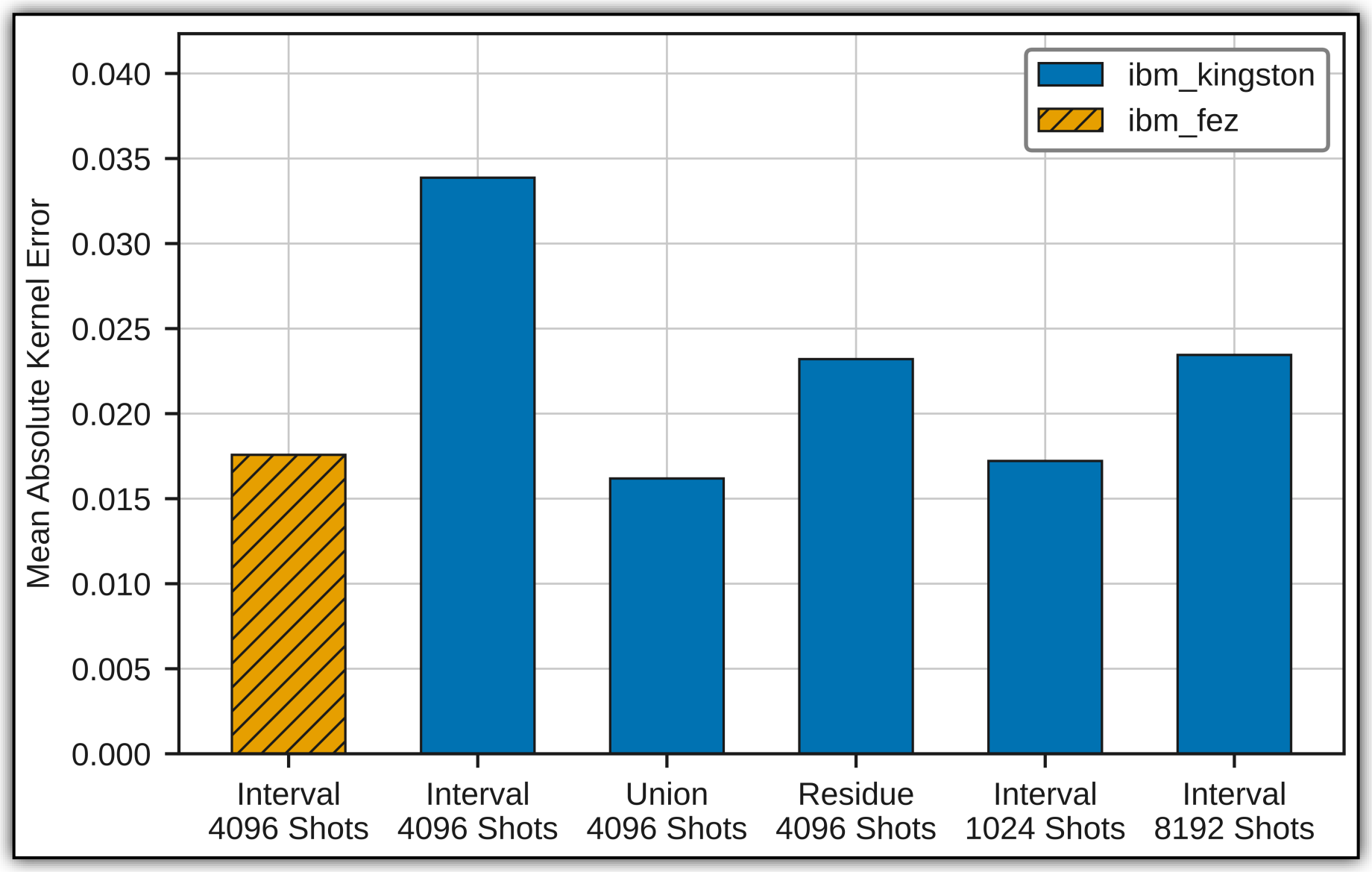

***Fig. 18*** *Mean absolute kernel error for each of the six hardware runs, spanning both processors and every predicate and shot budget tested*

One result here runs against the tidy story I expected to tell, and it is worth stating plainly instead of burying. More shots did not monotonically improve the agreement between the measured and the exact kernel. The 1024-shot interval run agreed slightly better than the 8192-shot run on the same device and the same predicate, which is the opposite of what shot noise alone predicts. The explanation is that at these budgets the device noise floor dominates the sampling noise, so run-to-run differences are set by calibration drift over the hours between submissions and not by how many shots were purchased. That is exactly why the clean inverse-square-root relationship belongs in the sampled-kernel simulation of Section 8.2, where the device is ideal, while the hardware numbers bury it completely. The hardware run establishes that the overlap circuit measures the right quantity and that the class structure survives real noise at this scale, while establishing nothing whatever about scaling, so I make no claim from it beyond feasibility. Work by Agliardi and colleagues, which carried covariant fidelity kernels to 156 qubits with a mitigation strategy built for them, is the reference point for what larger runs would require [34].

## 10. Discussion

The result to carry away is narrow and I would rather state it narrowly than let it inflate. A monitoring blind spot exists in fielded quantum key distribution that is provably hard for an efficient classical detector under a standard cryptographic assumption, and a quantum kernel whose feature map matches the adversary's predicate reads it. The simulation reproduces this at group sizes where classical simulation is faithful, with two seeds in agreement and a physics engine matching the decoy-state literature underneath. Three ingredients are load-bearing at once, which is unusual and worth noticing: remove the cryptographic structure and the quantum model loses its edge, remove the quantum kernel and nothing detects the pattern, remove the QKD setting and what remains is abstract learning theory with no operational consequence.

The stress tests are what turn a suggestive perfect score into a defensible claim, and they cost the headline number something. Once the exponent is recovered at finite precision with depolarising noise, and once the predicate is hardened past a single threshold, the quantum score drops in a measurable way while the classical monitor stays at chance. That descent is the honest picture of the advantage and it is more informative than the ceiling, because a deployment would operate on the descent. The hardware run adds a modest anchor underneath all of it, showing that the overlap circuit measures the kernel on a current device to within a few percent and that the classes still separate, which is the most a pre-fault-tolerant experiment can offer for a construction of this kind.

Read from the military side, the contribution amounts to assurance with no advocacy attached, because if an adversary can plant a cryptographically stealthy attack pattern on a command-and-control link, then a defender who predicts the attacked blocks can excise or down-weight exactly that key material, which converts a distributed and individually invisible attack into a recoverable one. The value holds whether or not quantum key distribution is the right key-agreement technology for a given deployment, and the authorities cited in Section 2.4 mostly think it is not. That disagreement does not dissolve the problem, because the links are being built regardless, and pattern-level monitoring is the kind of assurance a certified device does not provide on its own [1, 2, 16, 18].

## 11. Limitations

The limitations below are the ones that genuinely constrain the conclusions, and I have tried throughout to name the ones that sting instead of the ones that flatter. Each is stated so a reader can see which result it touches and how far that result would move if the assumption behind it failed. Several cut deep enough that I would not describe this work as deployable in anything like its present form.

The advantage is asymptotic and needs fault tolerance, since my detector recovers the discrete logarithm by brute force at small primes, which is an exact emulation of Shor's subroutine and is in no sense a quantum computer beating a classical one at scale. This paper demonstrates a mechanism and characterises its behaviour under degradation, together with a small check that the kernel circuit runs on hardware. It does not demonstrate, and cannot demonstrate, a hardware advantage at a size where the discrete logarithm is actually hard. Resource estimates for the fault-tolerant circuits involved still run to large numbers of physical qubits even after substantial improvement [48, 49, 50].

The classical hardness argument is conditional and, in its clean form, an evaluation-regime separation [35]. A fully general learning separation from textbook discrete logarithm hardness alone needs a stronger assumption than I invoke, and while the empirical collapse of the classical monitor is consistent with the claim, consistency is not proof and the theorem is scoped accordingly. Closing that gap is a question for learning theory rather than for the simulator, so I flag it as open instead of letting the measurements stand in for a result they cannot deliver.

The quantum kernel is only as good as the match between its feature map and the adversary's predicate, and the residue experiment shows how completely that can fail. A low-frequency map is blind to a Chinese-remainder pattern until the divisor frequencies are supplied. A real adversary is free to choose a predicate whose structure the defender's map does not contain, in which case the quantum detector inherits nothing, and designing maps that cover a broad family of predicates at once remains an open problem, well past any implementation detail. The concentration results of Thanasilp and colleagues are the general warning in this direction, since a poorly chosen quantum kernel washes out toward a constant and becomes untrainable as the system grows [56], and the mitigation work of Agliardi and colleagues shows the problem is live for covariant kernels specifically at utility scale [34]. My maps avoid that regime only because they are deliberately low dimensional and aligned to the label, which is the favourable case identified by Kübler and colleagues [31] and not a general property of the method.

The stealth assumption is credible for some primitives and not for others, because blinding and time-shift can both be tuned to stay inside the honest envelope, which is why the camouflaged datasets use blinding [4, 6, 7]. Photon-number-splitting is exposed by the decoy states and is a weak fit for hiding, and Trojan-horse leakage sits outside the monitored channel altogether [8, 9, 10]. The construction is strongest for the blinding-type case and I do not claim it covers every attack.

Every score reported here is a point estimate from a single run at a fixed seed, carrying no confidence interval. The two-seed protocol and the sweep across group sizes are what stand in for that, and they are weaker evidence than a distribution over repeated runs would be, particularly for the ablation, where the realistic row sits only about two hundredths below the idealised one on either seed. A reader who wants error bars would need to rerun the studies across many seeds, which the released code supports and this paper does not report. The telemetry underneath every simulation result in this paper is synthetic, produced by the model and never measured on a link. It comes from a physical model with realistic parameters, validated against the decoy-state key rate, but it is not data from a deployed device. A real link carries correlations and drifts the model omits, and whether a stealth footprint stays inside the envelope on real hardware is an empirical question this work does not settle. The key-rate model here is also asymptotic, whereas a deployment would use finite-key bounds, which are tighter [24].

## 12. Related Work

The closest prior work is the recent study of eavesdropping detection in BBM92 links by Gupta [57]. It builds a 24-dimensional temporal feature space over a 30-second sliding window on simulated free-space optical telemetry, characterising the shape and cross-observable correlation of the error rate, the Bell parameter, and coincidence rates. Evaluated over 24 hours of simulated data across five seeds, static error-rate thresholding reaches 17.3 percent recall against blended sub-threshold attacks while an XGBoost classifier on the engineered features reaches 96.9 percent recall at 97.6 percent precision. That work and this one agree that a naive threshold is not enough, and they differ on the nature of the attack, which is the whole point of comparison. Gupta's attacks are statistically detectable given the right features, which is why a classical model succeeds against them. My camouflaged class is constructed to be undetectable by any efficient classical detector under a cryptographic assumption, so no classical model can succeed and the separation is itself the result, which Table 12 sets out side by side.

**Table 12. Positioning against related detection work. The two result rows are not measured on one scale, since Gupta reports recall at a chosen operating point while I report area under the curve, so they are set side by side to contrast what each detector can reach rather than to rank them against each other.** Source: this work's column summarises out_seedA/benchmarks.json and out_seedB/benchmarks.json in the Ghillie archive [15], and the other column follows [57].

| Aspect | Classical machine learning for QKD, for example [57] | This work |
|---|---|---|
| Detector | Classical gradient boosting | Quantum kernel against classical baseline |
| Protocol and channel | BBM92 over free space | Decoy-state BB84 over fibre |
| Attack pattern | Statistically detectable | Classically undetectable under the discrete logarithm assumption |
| Classical monitor result | 96.9 percent recall | At chance in the hardness regime |
| Claim type | Empirical | Reduction, simulation, and a hardware feasibility check |
| Hardware needed for the effect | None | Fault tolerance, asymptotically |

Machine learning has been applied to quantum key distribution in several other ways, mostly with classical models, including predicting optimal decoy parameters, estimating key rates, and detecting attacks from side-channel features. These treat detection as an ordinary learning problem on features that carry the signal, and a classical learner is the right tool for that job. None of them place detection inside a concept class with a proven quantum-classical separation, which is the connection this paper draws.

On the learning-theory side the separation I use is that of Liu, Arunachalam and Temme [14] in the covariant-kernel form of Glick and colleagues [33], and it sits in a line of work on when quantum models can and cannot beat classical ones [11, 12, 13, 30]. Kübler and colleagues explain through the kernel spectrum why a low-dimensional, well-aligned kernel is the favourable case [31], Thanasilp and colleagues bound when a quantum kernel becomes untrainable at scale [56], and Agliardi and colleagues address both problems for covariant kernels on hardware [34]. Gyurik and Dunjko clarify the regime in which the separation holds cleanly, which is what bounds Section 4.4 [35]. To my knowledge this separation has been studied on abstract concept classes and has not been instantiated as a security-monitoring problem before.

On the attack side the primitives I model are standard, namely blinding [4, 5], time-shift [6, 7], photon-number-splitting [8], and Trojan-horse [9, 10], with the broader security picture in the reviews of Scarani and colleagues [25], Xu and colleagues [26], Pirandola and colleagues [27], and Lo, Curty and Tamaki [28]. What I contribute on this side is a way of hiding the pattern of an existing attack behind a cryptographic predicate, together with a detector that reads the pattern back out. The attack primitives themselves are left untouched, which is deliberate, because a construction that also required a new attack would be far easier to wave away as hypothetical.

## 13. Ethical and Dual-Use Considerations

This paper describes a defensive detector, and every attack primitive it uses is already published in the open literature. The camouflaged-attack construction arranges known attacks behind a known cryptographic predicate without introducing a new capability against a fielded system, and no operational attack code against a real device is released with it. Publishing is worthwhile because the blind spot exists whether or not anyone writes it down, and a defender who knows about it can build the monitoring layer that closes it. The work uses synthetic telemetry and a small benign kernel measurement on public cloud hardware. I cite primary sources throughout and mark clearly where a claim is conditional or asymptotic instead of demonstrated at scale. Readers who operate these links, which in practice means military and government operators, should treat the camouflaged-attack model as a reason to add pattern-level monitoring rather than as a reason to trust a certified device as a sealed guarantee.

## 14. Conclusion and Future Work

I have described a way to hide the pattern of a stealthy attack on a quantum key distribution link behind a discrete logarithm predicate, so that predicting which blocks were attacked is hard for an efficient classical monitor and tractable for a quantum kernel whose feature map matches the predicate. Across two seeds and group sizes from 6 to 14 bits, the classical monitor collapses to chance on the camouflaged class in the hardness regime while the quantum kernel separates it, and both detectors catch overt attacks. I then showed that the quantum advantage is not an artefact of a generous setup, by degrading the recovery to finite precision with depolarising noise and hardening the predicate, after which the quantum score falls to between 0.916 and 0.983 while the classical monitor stays at chance. The overlap circuit that measures a kernel entry runs on IBM Heron hardware and tracks the exact kernel to within 0.016 to 0.034 mean absolute error across six runs. Underneath all of it the physics engine reproduces a positive key rate to about 142 km.

Three directions follow from the point at which this work stops, and I list them in order of how soon each becomes tractable. The first is to move from a small-group hardware check toward larger and deeper kernels as devices improve, with an error-mitigated kernel estimate to narrow the gap against the exact values, for which zero-noise extrapolation and probabilistic error cancellation are the established starting points [58] and the covariant-kernel mitigation of Agliardi and colleagues is the closest precedent [34]. The second is to strengthen the reduction from the evaluation regime toward an identification-regime statement, and to widen the stealth analysis past the blinding-type primitive using measurements from a real device in place of a model. The third is the problem the residue predicate exposed, which is how to design feature maps covering a broad family of adversary predicates so the detector does not depend on guessing the adversary's structure in advance; the lattice-based version of the gate sits behind that same question and would tie the construction to a post-quantum assumption at the cost of a harder separation argument.

## Statements and Declarations

### CRediT Authorship Contribution Statement

Muhammad Shaheer Bin Junaid: Conceptualisation, Methodology, Software, Validation, Formal analysis, Investigation, Resources, Data curation, Writing (original draft), Writing (review and editing), Visualisation, Project administration.

### Declaration of Competing Interest

The author declares that he has no known competing financial interests or personal relationships that could have appeared to influence the work reported in this paper. This research received no specific grant from any funding agency in the public, commercial, or not-for-profit sectors.

### Data and Code Availability

The simulator, the classical and quantum detectors, and the datasets behind every table and figure are produced by the Ghillie package, which is released as open source under the MIT licence and archived with a permanent DOI at

https://doi.org/10.5281/zenodo.22906856 [15], with its development repository at https://github.com/mshaheerjunaid/Ghillie. Every run writes a timestamped benchmark record, and the sweep, sample-complexity, ablation, and feature-map alignment studies write their own records, so the numbers in this paper trace to specific runs and not to a re-execution that may have drifted since. Each hardware run writes a folder containing the measured kernel, the exact kernel, the backend name, the job identifier, and the calibration snapshot taken at execution time, and those six folders are included in the repository so that Tables 10 and 11 can be checked line by line against the raw output. The transpiled circuit diagrams exported from the IBM Quantum platform for all six jobs are archived alongside them, which is where the physical qubit assignments in Table 11 come from. The two seed configurations behind Tables 3 to 8, the alignment configuration behind Table 9, and the benchmark and hardware records are all part of the same release.